\documentclass[useAMS,usenatbib]{mn2e}

\usepackage{hyperref}
\usepackage{aecompl}
\usepackage{graphicx}
\usepackage{amssymb}
\usepackage{amsmath}
\usepackage{multicol}
\usepackage{longtable}
\usepackage{longtable}
\usepackage{float}
\usepackage{enumitem}
\usepackage{epstopdf}

\newcommand{\krome}{\textsc{Krome} }
\newcommand{\kromes}{\textsc{Krome}}

\title[Dust cooling and H$_2$ formation tables]{A detailed framework to incorporate dust in hydrodynamical simulations}

\author[T. Grassi et al.]{\parbox{\textwidth}{T. Grassi$^{1}$\thanks{Corresponding author: tgrassi@nbi.dk}, S. Bovino$^{2,4}$, T. Haugb\o lle$^{1,4}$, D. R. G. Schleicher$^{3}$}\vspace{0.4cm}\\
\parbox{\textwidth}{$^{1}$Centre for Star and Planet Formation, Niels Bohr Institute and Natural History Museum of Denmark, University of Copenhagen,
{\O}ster Voldgade 5-7, DK-1350 Copenhagen K, Denmark\\
$^{2}$Hamburger Sternwarte, Universit\"at Hamburg, Gojenbergsweg 112, 21029 Hamburg, Germany\\
$^{3}$Departamento de Astronom\'ia, Facultad Ciencias F\'isicas y Matem\'aticas, Universidad de Concepci\'on, Av. Esteban
Iturra s/n Barrio Universitario, Casilla 160, Concepci\'on, Chile\\
$^{4}$Kavli Institute for Theoretical Physics, University of California, Santa Barbara, California 93106, USA\\}}

\begin{document}

\newcommand{\enzo}{\textsc{Enzo} }
\newcommand{\enzos}{\textsc{Enzo}}
\newcommand{\flash}{\textsc{Flash} }
\newcommand{\flashs}{\textsc{Flash}}
\newcommand{\ramses}{\textsc{Ramses} }
\newcommand{\ramsess}{\textsc{Ramses}}
\newcommand{\fortran}{\textsc{Fortran} }
\newcommand{\python}{\textsc{Python} }
\newcommand{\pythons}{\textsc{Python}}
\newcommand{\dlsodes}{\textsc{DLSODES} }
\newcommand{\dlsodess}{\textsc{DLSODES}}
\newcommand{\ith}{$i$th }
\newcommand{\jth}{$j$th }
\newcommand{\nth}{$n$th }
\newcommand{\kth}{$k$th }

\newcommand{\dd}{\mathrm d}
\newcommand{\mA}{\mathrm A}
\newcommand{\mB}{\mathrm B}
\newcommand{\mC}{\mathrm C}
\newcommand{\mD}{\mathrm D}
\newcommand{\mE}{\mathrm E}
\newcommand{\mH}{\mathrm H}
\newcommand{\mHe}{\mathrm{He}}
\newcommand{\me}{\mathrm e}
\newcommand{\mSi}{\mathrm Si}
\newcommand{\mO}{\mathrm O}
\newcommand{\mX}{\mathrm X}
\newcommand{\cmc}{\mathrm{cm}^{-3}}
\newcommand{\real}{\mathbb R}
\newcommand{\superscript}[1]{\ensuremath{^{\scriptscriptstyle\textrm{#1}\,}}}
\newcommand{\trader}{\superscript{\textregistered}}
\newcommand{\eqn}[1]{Eqn.(\ref{#1})}
\newcommand{\eqnrange}[2]{Eqns.(\ref{#1}) to (\ref{#2})}

\newcommand{\arl}[1]{\url{#1}}
\newcommand\mnras{MNRAS}
\newcommand\apj{ApJ}
\newcommand\apjs{ApJS}
\newcommand\apjl{ApJL}
\newcommand\aap{A\&A}
\newcommand\apss{Ap\&SS}
\newcommand\ssr{Springer}
\newcommand\jgr{JGeophysRes}
\newcommand\icarus{Icarus}

\renewcommand\arraystretch{1.7}

\newcommand{\tgcomment}[1]{{#1}}

\newcommand\osu{osu\_01\_2007}
\def\tm{\leavevmode\hbox{$\rm {}^{TM}$}}

\date{Accepted *****. Received *****; in original form ******}

\pagerange{} \pubyear{2012}

\maketitle

\label{firstpage}

\begin{abstract}
Dust plays a key role in the evolution of the ISM and its correct modelling in numerical simulations is therefore fundamental. We present a new and self-consistent model that treats grain thermal coupling with the gas, radiation balance, and surface chemistry for molecular hydrogen self-consistently. This method can be applied to any dust distribution with an arbitrary number of grain types without affecting the overall computational cost.
In this paper we describe in detail the physics and the algorithm behind our approach, and in order to test the methodology, we present some examples of astrophysical interest, namely (i) a one-zone collapse with complete gas chemistry and thermochemical processes, (ii) a 3D model of a low-metallicity collapse of a minihalo starting from cosmological initial conditions, and (iii) a turbulent molecular cloud with H-C-O chemistry (277 reactions), together with self-consistent cooling and heating solved on the fly. Although these examples employ the publicly available code \kromes, our approach can easily be integrated into any computational framework.
\end{abstract}

\begin{keywords}
astrochemistry -- ISM: evolution, dust -- methods: numerical.
\end{keywords}

\section{Introduction}
The microphysics of dust grains plays a key role in many astrophysical environments, especially in star-forming regions during the gravitational collapse, where the thermal history of the gas is dominated by the gas cooling due to gas-grain collisions and heating from the formation of molecular hydrogen on the dust surface (e.g. \citealt{Hollenbach1979,Hollenbach1989,Omukai2005,Schneider2006,Tsuribe2006}).

The collisions between gas particles and dust grains result in a net gas cooling\footnote{Note that this actually becomes heating when the gas temperature is lower than the dust temperature, but we always employ the term ``cooling'', unless otherwise stated.} and dust grain heating that is balanced by the radiation emitted from the grain surface, and absorbed from external sources, e.g.~nearby stars or the Cosmic Microwave Background (CMB) radiation \citep{Hollenbach1979}. Additionally, the opacity of the dust grains and their ability to reprocess radiation in to the far infrared has a considerable impact on the radiation field, and for this reason it should be included consistently when the dust physics is modelled \citep{Semenov2003,Omukai2005}. Finally, the dust also catalyses the formation of molecules, most importantly H$_2$, but molecule formation on grain surfaces depends sensitively on the surface temperature of the grains \citep{Cazaux2009,Hocuk2015}.

Unfortunately, from a numerical point of view dust requires detailed, complex, and computational expensive modelling (e.g.~see \citealt{Budaj2015,Camps2015,Mattsson2015}). Large-scale simulations therefore often rely on a simplified description of the dust microphysics \citep{Dopcke2011,Meece2014,Yajima2014}, e.g. assuming averaged dust properties, approximated H$_2$ formation reaction rates, constant dust temperature, or ignoring any interaction with the radiation field. Due to its complexity, dust has been included with a detailed treatment only in a few 3D models. In particular, \citet{Dopcke2011,Dopcke2013}, \citet{Meece2014}, and \citet{Smith2015} use  a similar methodology based on average dust properties following \citet{Hollenbach1979,Hollenbach1989} and adopt a piecewise polynomial approximation for the grain opacity. However, their models implicitly assume power-law distributions and averaged optical properties, which might be inaccurate in some environments, as recently shown by \citet{Bovino2016}. Other works include accurate dust formation and destruction with adaptive mesh refinement codes (e.g.~\citealt{McKinnon2016}), and H$_2$ formation on grain surfaces \citep{Bekki2013}, but do not treat the gas-radiation-dust thermal balance and associated cooling.
On the other hand, \citet{Chiaki2016} have proposed a model with detailed dust physics, including grain growth, but as shown in our tests in Sect.~\ref{sect:onezone} and in \citet{Bovino2016} a detailed account of the dust physics tends to be expensive from a numerical point of view compared to our new table-based approach, which can be decisive for CPU time limited models with a large amount of computational elements.

To reduce the computational cost we here present a light-weight approach based on tables that allows accurate calculations even in 3D hydrodynamical models. Even though it is table-based, it is fully-consistent and can be employed in numerical simulations at a very low cost to obtain on the fly (i) averaged dust temperature,  (ii) molecular hydrogen formation rates, and (iii) cooling using simple bilinear or trilinear interpolation on a regular grid. \tgcomment{All the presented tests use the publicly available code \kromes\footnote{\url{http://kromepackage.org}} \citep{Grassi2014} to model chemistry and microphysics, and to embed dust tables. We however note that the tables can be included in any framework by using a generic interpolation routine.
The tables employed in this paper and in \citet{Bovino2016} are available on the \krome website\footnote{\url{http://kromepackage.org/dust_tables/}}.}

In Sect.~\ref{sect:Tdust_and_cooling} we describe the main equations for dust temperature and cooling, and in Sect.~\ref{sect:H2} the formation of H$_2$ on grains. In Sects.~\ref{sect:onezone}-\ref{sect:hydro_cloud} we present the applications: a one-zone collapse, 3D collapse of a minihalo simulated with \enzos, and a 3D model of the magnetised ISM simulated with \ramsess.

\section{Dust temperature and cooling}\label{sect:Tdust_and_cooling}
In this Section we discuss how to compute dust cooling and grain temperatures for first an optically thin medium and then for optically thick environments.
\subsection{Optically thin medium}
In temperature equilibrium a single dust grain is regulated according to Kirchhoff's law: the radiation absorbed ($\Gamma_{abs}$) by the grain is equal to the emitted one ($\Gamma_{em}$), and we extend this thermal balance with the gas cooling function due to dust ($\Lambda$), in order to have
\begin{equation}\label{eqn:rad_balance_simple}
 \Gamma_{em} = \Gamma_{abs}+\Lambda\,.
\end{equation}
Integrating over a distribution of dust grains embedded in a gas in the \emph{optically thin regime} this becomes \citep{TielensBook}
\begin{eqnarray}\label{eqn:rad_balance}
 &&\int_{a_{min}}^{a_{max}}\pi a^2 \varphi(a) \int_{0}^{\infty} \frac{Q_{abs}(a,E) B\left[E,T_d(a)\right]}{h} \dd E\, \dd a\nonumber\\
 &=& \int_{a_{min}}^{a_{max}}\pi a^2 \varphi(a) \int_{0}^{\infty} \frac{Q_{abs}(a,E) [J(E)+J_z(E)]}{h} \dd E\, \dd a\nonumber\\
 &+& 2 f v_g n_{tot}\int_{a_{min}}^{a_{max}}\pi a^2\varphi(a) k_B\left[T_g-T_d(a)\right] \dd a\,,
\end{eqnarray}
where $v_g$ is the thermal speed of the gas, $\pi a^2$ the geometrical cross section of a grain of radius $a$, 
$Q_{abs}(a,E)$ the absorption coefficient for the given grain material, $B\left[E,T_d(a)\right]$ the spectral radiance of a black-body with a given grain temperature $T_d(a)$, $J(E)$ the impinging flux on the dust grain, $k_{B}$ the Boltzmann constant, and $J_z(E)$ the CMB radiation at redshift $z$. The factor $f$ accounts for gas-grain collisions with atoms and molecules other than atomic hydrogen, such as He and charged species \citep{Hollenbach1979,Schneider2006}, and it depends not only on the gas composition, but also on the grain charge distribution (see \citealt{Draine2009} sect.~24.1.2).  For the presented applications we use a constant  $f=0.5$ to mimic a partially molecular cold gas \citep{Hollenbach1979}, and we postpone a more complete modelling of $f$ to future work. The grain size distribution $\varphi(a)$ is defined in the range $a_{min}$ to $a_{max}$, and is normalised by the dust mass density
\begin{equation}\label{eqn:normalization}
 \rho_{d} = D_\odot\,10^Z\,n_{tot}\,\mu_g\,m_p = \mathcal{C} \frac{4}{3}\pi\rho_0\int_{a_{min}}^{a_{max}}a^3\varphi(a)\dd a\,,
\end{equation}
with $\rho_d$ the total dust mass density, $D_\odot$ the dust-to-gas mass ratio at solar metallicity in the Milky Way, $Z$ the metallicity ($Z=0$ is solar), $n_{tot}$ the total gas number density, $\mu_g$ the mean gas molecular weight, $m_p$ the mass of the proton, $\mathcal{C}$ the normalization constant, and $\rho_0$ the bulk density of the grain. We have assumed that the dust-to-gas mass ratio depends linearly on the metallicity $D=D_\odot10^Z=\rho_d/\rho_g$. However, for high redshift applications this linearity is no longer valid (e.g.~see \citealt{Schneider2012,Nozawa2006}). The term $\mathcal{C}\varphi(a)$ corresponds to the number density per grain size, $\dd n(a)/\dd a$. \eqn{eqn:normalization} is used to find $\mathcal{C}$. The equation can be extended to a mix of grain types assuming that the mass density of the \jth grain type is related to the total dust mass as $\rho_d = \sum_{j}\rho_{d,j}x_j$, where $x_j$ is a scaling factor that depends on the relative abundance, which we assume to be solar, of the \jth key element and satisfies $\sum_{j}x_j=1$. In order to keep the notation simple the equations presented in this paper are written for a single grain type only, however in the numerical implementation we are using multi-species equations, including in \eqn{eqn:rad_balance} and \eqn{eqn:normalization}, to compute the dust temperature profile $T_d(a)$.

The size distributions and grain compositions are not constrained and can be adapted to take on any form depending on the environment (e.g.~interstellar medium and SNe-originated dust), and throughout this paper we make no assumptions about the $\varphi_j(a)$ functions. This allows us to compute the dust cooling, molecular hydrogen formation on grains, and dust temperature for the specific application instead of using averaged properties, such as in e.g.~\citet{Hollenbach1979} where a standard galactic size distribution and composition is assumed.

\subsubsection{Binned distribution}\label{sect:physics_bin_notation}
To discretise the dust distribution function we subdivide it in to $N_d$ bins spaced logarithmically in size. The number density in the \ith bin is
\begin{equation}
n_{d,i} = \mathcal{C}\int^{(a_i a_{i+1})^{1/2}}_{(a_{i-1} a_i)^{1/2}} \varphi(a) \dd a\,,
\end{equation}
where the integration limits are at the logarithmic midpoints. $\varphi(a)$ is then a step function and values are well represented at the midpoint, making it possible to compute the outer integrals in \eqn{eqn:rad_balance} as sums. If the distribution $\varphi$ is known analytically, more accurate integration techniques are possible (see Sect.~\ref{sect:param_sensitivity}.). However, since we aim at keeping the method as general as possible, we will in general use a large number of bins $N_d$, for which the discrete distribution will approximate well a smooth distribution.
The dust temperature is a quantity that, \emph{in the optically thin regime}, can be computed for a single bin, since in this regime each grain can be considered individually. Hence, we can remove the outer integral over the distribution and for a grain in the \ith bin we have
\begin{eqnarray}\label{eqn:rad_balance_bin}
 &&\int_{0}^{\infty} \frac{Q_{abs}(a_i,E) B\left[E,T_{d,i}\right]}{h} \dd E \nonumber\\
 &=& \int_{0}^{\infty} \frac{Q_{abs}(a_i,E) [J(E)+J_{z}(E)]}{h} \dd E \nonumber\\
 &+& 2 f v_g n_{tot} k_B\left[T_g-T_{d,i}\right]\,,
\end{eqnarray}
which is a transcendental equation that can be solved numerically (e.g. with a bisection method) in order to find the unknown $T_{d,i}$. For a given impinging radiation field $J(E)$, the resulting temperature \emph{in a single bin} is then a function of $T_g$ and $n_{tot}$ only.

The contribution from the \ith dust bin to the dust cooling is then
\begin{equation}\label{eqn:partial_cooling}
 L_i = \pi a_i^2 n_{d,i} k_B\left[T_g-T_{d,i}\right]\,.
\end{equation}
\eqn{eqn:partial_cooling} can be integrated over the bin distribution to find the total gas cooling
\begin{equation}\label{eqn:total_cooling_bin}
 \Lambda = 2 f v_g n_{tot} \sum_{i=1}^{N_d} L_i\,.
\end{equation}

Our method uses a size-dependent dust temperature, while in most applications only a single dust temperature is used (see Sect.~\ref{sect:onezone}). We could define a dimensionally averaged dust temperature
\begin{equation}
 \langle T_{d}\rangle^\ell = \frac{\int_{a_{min}}^{a_{max}}T_{d}(a)\varphi(a)a^\ell\dd a}{\int_{a_{min}}^{a_{max}}\varphi(a)a^\ell\dd a}\,,
\end{equation}
weighted according to number density ($\ell=0$), area ($\ell=2$), or mass ($\ell=3$) and discretised analogously to \eqn{eqn:total_cooling_bin}, but the dust temperature is determined by the radiative balance. Therefore a better alternative to dimensional weighting is to use the spectral radiance of the grain population to find the black body equivalent temperature $\langle T_d \rangle$ by solving
\begin{eqnarray}
 && \int_E \langle Q_{abs}(E)\rangle_a B\left[E,\langle T_d\rangle\right]\dd E \nonumber\\
        &=&\frac{\int_a \varphi(a)a^2 \int_E Q_{abs}(a,E) B\left[E,T_d(a)\right]\dd E\,\dd a}{\int_a\varphi(a)a^2\dd a}\,,
\end{eqnarray}
where the absorption coefficient is weighted according to the grain area
\begin{equation}
 \langle Q_{abs}(E)\rangle_a = \frac{\int_a \varphi(a)a^2 Q_{abs}(a,E)\dd a}{\int_a \varphi(a)a^2 \dd a}\,.
\end{equation}
This has the additional advantage that at high densities where gas and dust is tightly coupled, all the bins have $T_{d,i}=T_{gas}\,\,\forall\,i$, and $\langle T_d\rangle$ becomes the real dust temperature. We therefore recommend using $\langle T_d\rangle$ as a representative temperature.

\subsection{Optically thick medium}\label{sect:thick}
In the optically thick regime the impinging radiation field changes depending on the environment and the optical depth. For simplicity, to keep the cooling and the averaged temperature as a function of $T_g$ and $n_{tot}$ only, we here discuss two representative cases, namely (i) a molecular cloud-like ISM and (ii) a high-density collapsing cloud. By extending the tables to depend on visual extinction or the intensity in key frequency bands, the method would be applicable in the case of a general radiation field, as long as it is parameterised with a reasonable number of parameters (see Sect.~\ref{sect:onezone} for additional details). These two cases illustrates two limits: In the first limit -- case (i) -- the optical depth does not change appreciably inside a cell, or the change in optical depth is only due to the gas. In that case each dust size bin respond independently to the radiation, given the local radiation field, though the total optical depth may depend non-trivially on the integrated dust opacity. In the second limit, the dominating optical depth is inside a single cell, and therefore to compute the energy balance \eqn{eqn:rad_balance_simple} it has to be taken in to account that not all photons can escape through the optically thick cell. This is a truly optically thick problem, and exemplified by case (ii).

\subsubsection{Molecular cloud ISM}\label{subsub:ism}
We assume for this case a global external (environment-dependent) radiation field and an attenuation method that is a function of the local properties of the gas. If the unattenuated radiation spectrum is $J_0(E)$, the impinging radiation field is
\begin{equation}
 J(E) = J_0(E) \,e^{-\tau_v}\,,
\end{equation}
where $\tau_v = 0.9208 A_v$.
In the present paper we use the following approximation for the visual extinction that depends only on the gas density $n_{tot}$
\begin{equation}
 A_{v,ISM} = \left(\frac{n_{tot}}{10^{3}\,{\rm cm^{3}}}\right)^{\alpha}\,,
\end{equation}
where $\alpha=2/3$. We have made this empirical relation by fitting a power-law above number densities of
200 cm$^{-3}$ to the density-$A_v$ relation from the hydro-chemical models including full radiative transfer by \citet{Glover2010}. See also the discussion in \citet{Safranek2016}.
We model the external radiation as a background Draine's flux, defined as
\begin{eqnarray}
 J_{0,ISM} &=& h \left(\frac{E}{\rm eV}\right) \left[1.658\times10^6\left(\frac{E}{\rm eV}\right)\right.\nonumber\\
 &-& 2.152\times10^5 \left(\frac{E}{\rm eV}\right)^2\nonumber\\ 
 &+& \left.6.919\times10^3\left(\frac{E}{\rm eV}\right)^3\right]  \,\,{\rm erg\,cm^{-2}\,sr^{-1}}\,,
\end{eqnarray}
which is valid in the range $E=[5, 13.6]$~eV.
This model is applied in Sect.~\ref{sect:hydro_cloud}.

\subsubsection{High-density collapsing cloud}\label{subsub:besc}
In a collapsing cloud at high enough densities (more than $10^{12}$ to $10^{18}$~cm$^{-3}$ depending on the metallicity), the radiation emitted and absorbed is affected not only by the gas opacity, but also the presence of the dust itself.

We model the opacity inside a single computational element by introducing (e.g. \citealt{Omukai2005}) an escape probability $\beta_e$ in \eqn{eqn:rad_balance_simple}
\begin{equation}\label{eqn:beta_balance_basic}
\beta_e\,\left[\Gamma_{em}-\Gamma_{abs}\right] =  \Lambda\,.
\end{equation}

This equation can be rewritten for a single grain in the \ith bin as
\begin{equation}\label{eqn:beta_balance}
 \beta_e(\mathbf{T_d})[\Gamma_{em}(T_{d,i}) - \Gamma_{abs}] = \Lambda(T_{d,i})\,,
\end{equation}
where $\mathbf{T_d}=\{T_{d,i}\,\forall i\}$ represents the set containing the temperatures of all the bins. This is necessary because in the multiple bins approach a grain is not affected only by the opacity coming from the grains with the same properties, but by all grains (i.e. with different temperatures, optical properties, and compositions).

The escape probability is calculated as in \citet{Omukai2005}
\begin{equation}
 \beta_e = \min\left[1, \,(\tau_g+\tau_d)^{-2}\right]\,,
\end{equation}
where the dust opacity is given by
\begin{equation}\label{eqn:dust_opacity}
 \tau_d = l_J\pi\sum_i\,n_{d,i}\,a_{i}^2 \frac{\int_0^\infty Q_{abs}(E,a_i) B(E,T_{d,i})\dd E}{\int_0^\infty B(E,T_{d,i}) \dd E}\,,
\end{equation}
with $n_{d,i}$ the dust number density in the \ith bin, $a_{i}$ the grain size, and the Jeans length $l_J$
\begin{equation}\label{eqn:jeans}
 l_J = \sqrt\frac{\pi k_B T_g}{\rho_g m_p G \mu_g}\,,
\end{equation}
where $m_p$ is the proton mass, $G$ the gravitational constant, and $\mu_g$ the mean molecular weight.

In the optically thick regime, \eqn{eqn:beta_balance} depends not only on the temperature of the \ith grain $T_{d,i}$, but also on the temperatures of all the other dust bins $\mathbf{T_d}$, through the sum in \eqn{eqn:dust_opacity}. For this reason the bisection method usually applied to \eqn{eqn:beta_balance} is not sufficient to find the dust temperature, but one should solve a non-liner system of $N_d$ \eqn{eqn:beta_balance} in order to find the roots $\mathbf{T_d}$ that represent the set of the grain temperatures. We solve it by computing \eqn{eqn:beta_balance} for all the dust bins iteratively until convergence (see Appendix~\ref{sect:convergence_dust}).

Even though $l_J$ is a function of the mean molecular weight $\mu_g$, which depends on the chemical composition of the gas, for a given $n_{tot}$ both $\rho_g$ and $n_{d,i}$ are linear functions of $\mu_g$ (in fact, $D\,\rho_g=D\,n_{g}\mu_g m_p=\rho_d=4/3\pi\rho_0\sum_i n_{d,i}\langle a_i\rangle^3$) and $\tau_d$ does not depend on the mean molecular weight. Therefore the lookup tables do not have any dependence on the detailed gas composition.

The corresponding gas opacity $\tau_g$ is obtained from the results of \citet{Mayer2005}, where in their Tab.~E2 they show the Planck opacity for a gas with POP~III composition (their Tab.~1). Their data are valid in the range $60\lesssim T_g\lesssim 4\times10^4$~K and $10^{-16}\leq\rho_g\leq10^{-2}$~g~cm$^{-3}$. \tgcomment{The POP~III composition included here follows the implementation that can be found in \kromes, but different environments might require different gas opacities. We note that for the applications pursued here, the gas opacities are always sub-dominant as compared to the dust opacities, and we expect that the inclusion of metals would not make a difference. We however note that in the high-energy regime (e.g. X-rays), it will be important to include realistic opacities for the metals as well (e.g. \citealt{Meijerink2005}).}

This model is applied in Sects.~\ref{sect:onezone} and~\ref{sect:hydro}.
\subsection{Dust evaporation}\label{subsub:evaporation}
In our table-based approach we have to assume that the dust distribution is exclusively a function of the local gas conditions, it cannot have a history. The easiest zeroth order approach is to use a constant dust distribution, but at high dust temperatures the dust grains will be destroyed due to evaporation of molecules and atoms from the surface. This process is controlled by the binding energy of the atom to the surface lattice. It can be modelled by assuming the atom is attached with a spring to the lattice with a frequency $\nu_0$, which is the bulk Debye frequency $\nu_0=\Theta_Dk_B/h$, where $\Theta_D$ is the Debye temperature and $h$ the Planck constant. The probability $p_l$ and the time $t_l$ in which \emph{a layer} detaches from a dust surface with a temperature $T_d$ is then described by the Polanyi-Wigner equation \citep{Stahler1981}
\begin{equation}\label{eqn:evap_prob}
	p_l = \frac{1}{t_l} = \nu_0\exp\left(-\frac{E_0}{k_BT_d}\right)\,,
\end{equation}
where $E_0$ is the binding energy of the dust. The thickness of the layer is $\Delta a \approx \left(m_j/\rho_0\right)^{1/3}$, where $m_j$ is the  mass of the atoms in the shell with bulk density $\rho_0$. If we assume that the grain composition is homogeneous, i.e. with constant $E_0$ and $\nu_0$ as a function of $a$, we can integrate \eqn{eqn:evap_prob} to find the evaporation time of a grain. This is the time required to remove \emph{all layers} of a grain \citep{Draine2009}
\begin{equation}\label{eqn:evaporation}
	t_e = \frac{a_0}{\nu_0\,\Delta a}\exp\left(\frac{E_0}{k_BT_d}\right)\,.
\end{equation}

\subsubsection{Dust evaporation during cloud collapse}
\begin{figure}
  \centering
  \includegraphics[width=.49\textwidth]{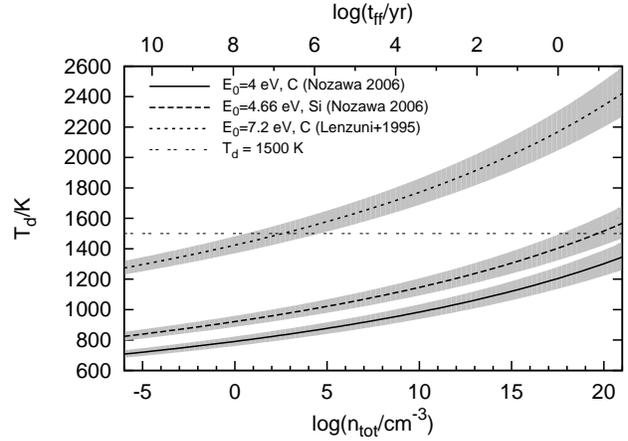} 

  \caption{Evaporation temperature as a function of the gas density assuming $t_e = t_{ff}$. Solid lines represent  a grain model with $a_0=10^{-6}$~cm, $\rho_0=2.25$~g~cm$^{-3}$, $\nu_0=10^{12}$~s$^{-1}$ and $m_j=12\,m_p$. Different lines are for different binding energies $E_0$ (see legend). The grey areas represent an increase/decrease by one order in magnitude of $a_0$. Note that increasing the size also increases the required dust temperature $T_d$. The horizontal line indicates $T_d=1500$~K as reference.}
  \label{fig:evap}
 \end{figure}

In the previous Section we found that, once the grain parameters are set, the evaporation time $t_e$ is a function of the dust temperature $T_{d,i}$ only. The evaporation time should be compared with the typical time-scale of the problem, which in a cloud collapse is the free-fall time
\begin{equation}\label{eqn:freefall}
	t_{ff} = \sqrt{\frac{3\pi}{32 G\rho_g}}\,,
\end{equation}
where $G$ is the gravitational constant and $\rho_g=n_{tot}\mu_g m_p$ is the mass density of the gas. The condition for evaporation is then $t_e\le t_{ff}$.
To test the dust behaviour during the collapse we assume a default set of carbon-like grain parameters, namely $a_0=10^{-6}$~cm, $\rho_0=2.25$~g~cm$^{-3}$ \citep{Zhukovska2008}, $\nu_0=10^{12}$~s$^{-1}$ \citep{Draine2009}, and $m_j=12\,m_p$. In Fig.~\ref{fig:evap} the solid lines represent the evaporation temperature as a function of $n_{tot}$ (assuming $\mu_g=1.22$) when $t_{ff}=t_{e}$, i.e.~the minimum temperature required to completely destroy a grain with the previously mentioned characteristics. The lines are for different binding energies as in the legend, namely (i) carbon grains with $E_0=4$~eV, (ii) silicon grains with $E_0=4.66$~eV  \citep{Nozawa2006}, and  (iii) carbon grains with  $E_0=7.2$~eV \citep{Lenzuni1995}.
The dependence on $n_{tot}$ arises from the definition of free-fall in \eqn{eqn:freefall} when coupled to \eqn{eqn:evaporation}. As expected the evaporation temperature grows with the gas density, because the time-scale of the problem is shorter, and a faster evaporation mechanism (i.e. higher temperature) is required. In the same figure we increase and decrease the size $a_0$ in \eqn{eqn:evaporation} by one order of magnitude (grey areas around the solid curves). A larger grain implies a longer evaporation time, so that a higher $T_d$ is required to satisfy the condition $t_{ff}=t_{e}$. 
As expected, the binding energy, which is also the largest unknown \citep{Draine2009}, affects the results dramatically when compared to the effects of the size variation.

The temperature function shown in Fig.~\ref{fig:evap}  does not take into account possible sticking from the gas, and assumes no thermal history of the grain that will experience partial evaporation due to lower temperature values. The latter implies that a time-dependent size evolution is required to properly model evaporation. However, the steep dependence of the evaporation time with the dust temperature shown by \eqn{eqn:evaporation} suggests that the process becomes important only when the dust temperature is close to the critical temperature, hence sticking will not be able to compete with evaporation, and neglecting the grain thermal history is a reasonable approximation.

We have included this instantaneous dust evaporation effect in our tables, assuming that the free-fall collapse time is the relevant time-scale in the problem. This lowers the density of the dust grains, as a function of size, and translates in to a lower cooling rate and H$_2$ formation efficiency at high dust temperatures. We have modelled dust evaporation as a size-independent process, so that all the grains of a given type are affected in the same way. This approximation is justified by the results reported in Fig.~\ref{fig:evap}. On the other hand, this process has a strong dependence on the grain material: it could selectively reduce the opacity (by destroying grains with a lower binding energy) leaving only grains with higher binding energies available for cooling and H$_2$ formation. However, the temperatures in the tests reported in this paper are not high enough to show this effect.

\section{Molecular hydrogen formation on grains}\label{sect:H2}
In the previous sections we introduced the formalism to determine the dust temperature of each bin. Given dust and gas temperatures, we can determine the molecular hydrogen catalysis following \citet{Cazaux2009}
\begin{equation}\label{eqn:H2formation}
 \frac{\dd n_{\rm H_2}}{\dd t} = \frac{n_{\rm H}v_g}{2}\,\mathcal{C}\int_{a_{min}}^{a_{max}}\pi a^2\,S\left[{T_g,T_d(a)}\right]\epsilon\left[{T_g,T_d(a)}\right] \varphi(a) \dd a\,.
\end{equation}
The dependence on dust temperature $T_d$ in \eqn{eqn:H2formation} is given by the sticking coefficient ($S$) and the efficiency factor ($\epsilon$). The former is defined as
\begin{equation}\label{eqn:sticking}
 S\left[{T_g,T_d(a)}\right] =\left[1+0.4\,\sqrt{T_2+\frac{T_d(a)}{100\,\rm K}} + 0.08\,T_2^2\right]^{-1}\,,
\end{equation}
where $T_2=T_g/ 100{\,\rm K}$, while the latter depends on the \jth grain material. See Appendix~\ref{appendix:epsilon} for the efficiency factor of carbonaceous and
silicate grains. Note that new experiments on sticking have recently been discussed by \citet{He2016}, but for consistency with \citet{Cazaux2009} we use \eqn{eqn:sticking}.

\subsection{Binned distribution}
If we consider  $N_d$ size-bins for the distribution, \eqn{eqn:H2formation} for the \ith bin leads to the following expression
\begin{equation}\label{eqn:H2formation_bin}
 F_{i} = \pi a_{i}^2S_{i}\epsilon_{i} n_{d,i}\,,
\end{equation}
and the H$_2$ formation rate for the total distribution is
\begin{equation}\label{eqn:H2formation_bin_tot}
 \frac{\dd n_{\rm H_2}}{\dd t} = \frac{n_{\rm H}v_g}{2}\,\sum_{i=1}^{N_d}F_{i}\,.
\end{equation}
This applies to $N_d$ bins of a single type of dust, and should in general be summed over all the grain types. Moreover, since \citet{Cazaux2009} only consider carbon- and silicon-based grains, we assume that every type of dust (except amorphous carbon) forms H$_2$ as silicon-based grains. Just as with the dust temperature, the expressions for H$_2$ formation can be refined given an analytical grain distribution.

\section{Application 1: one-zone low-metallicity cloud collapse with dust}\label{sect:onezone}
In Sect.\ref{subsub:besc} we found that the main parameters needed
to describe the cooling and H$_2$ formation rate are the gas temperature ($T_g$) and density ($n_{tot}$), the type of radiation $J(E)$, and the dust-to-gas ratio ($D$). For the purposes of this application we freeze some of the parameters, since the simulated environment here is a low-metallicity cloud. We use a set-up similar to \citet{Omukai2005}, including (i) cooling from  H$_2$ \citep{Glover2015}, Compton, CI, CII, OI, OII, continuum, and from endothermic reactions, (ii)  heating from exothermic reactions (including H$_2$ on dust) and from gas compression, (iii) H$_2$ opacity, (iv) an adiabatic index consistent with chemistry, and (v) a chemical network with 75~species (i.e.~\verb+react_primordialZ+ of \kromes). More details can be found in \citet{Grassi2014} and \citet{Omukai2005}. For this test the only on-the-fly parameters (i.e. parameters that change during the calculation) are $T_g$ and $n_{tot}$. We assume constant $D_\odot=0.00934$ (that is rescaled linearly with the metallicity $D=D_\odot\,10^Z$), and a CMB radiation field with $T_{CMB}=2.73(1+z)$ at $z=16$. The opacity is calculated using the $\beta_e$ term described in Sect.~\ref{subsub:besc}. For simplicity we assume that the opacity is dominated by the dust and $\tau_g=0$ (e.g.~see \citealt{Semenov2003}).

Under the previous assumptions we define three functions, namely $f_{\rm H_2}$ for molecular hydrogen formation, $f_{\Lambda}$ for cooling, and $f_{\langle T_d\rangle}$ for averaged dust temperature that encapsulate the processes described in the previous sections. They are employed in our calculations in the following way (see \citealt{Grassi2014} for more details)
\begin{eqnarray}
 \dot n_{\rm H_2} &=& \mu_g\,n_\mH\,f_{\rm H_2}\mkern-6mu\left(T_g, n_{tot}\right)n_{tot}\label{eqn:fH2}\\
 \Lambda &=& \mu_g\,f_{\Lambda}\mkern-6mu \left(T_g, n_{tot}\right)n_{tot}^2\label{eqn:fCool}\\
 \langle T_d\rangle &=& f_{\langle T_d\rangle}\mkern-6mu\left(T_g, n_{tot}\right)\label{eqn:fTdust}\,,
\end{eqnarray}
where the first two equations are required during the
calculation. In our approach the complete dependence on the dust
temperature is factored out and hidden in the tables,
and used self consistently for cooling and H$_2$ formation,
and the third table is only used as a post-processed
diagnostic.
The functions are represented by three look-up tables computed using the formalism described in the previous sections. We have factored out the number density
dependence, $n_{tot}$ and $n_{tot}^2$, explicitly in \eqn{eqn:fH2} and \eqn{eqn:fCool} to improve the quality of the fit by removing the span in orders of magnitude between different densities. 
Furthermore, we can generate individual tables for the dust temperature of each bin (i.e. $f_{ T_{d,i}}$) when this is required for particular purposes, e.g. bin-by-bin surface chemistry other than H$_2$ catalysis.

In this test we assume a MRN distribution \citep{Mathis1977} $\varphi(a)\propto a^{-3.5}$, in the range $a_{min}=5\times 10^{-7}$~cm, $a_{max}=2.5\times 10^{-5}$~cm, made of carbonaceous and silicate grains with bulk density $\rho_0=2.25$~g~cm$^{-3}$ and $\rho_0=3.13$~g~cm$^{-3}$, respectively \citep{Zhukovska2008}. The molecular hydrogen catalysis parameters are from \citet{Cazaux2009} as described in Sect.~\ref{sect:H2}, while for the absorption coefficient $Q_{abs}$ we use the optical properties integrated in the energy range $1.32\times10^{-3}\le E\le1.24\times10^3$~eV (see Appendix~\ref{appendix:qabs}). Note that the computational cost of the tables is not affected by the presence of different dust types with different optical properties and by the number of energy bins for the impinging radiation field, so that we can assume arbitrary precision in the integrals in \eqn{eqn:rad_balance}. The tables are
produced using fully-consistent calculations described in the previous sections. Moreover, even if in this example we employed a MRN distribution, the tables can be generated for any arbitrary $\varphi(a)$, as e.g. in \citet{Bovino2016}.

The tables span the following parameter domain $2\le T_g\le 10^4$~K and $10^{-2}\le n_{tot}\le 10^{18}$~cm$^{-3}$, with a $50\times50$ logarithmically spaced grid, which is interpolated using a standard bilinear method. The number of grid points is determined from convergence, and since the tables fit entirely in cache, the exact number of points does not affect the computational efficiency.

We use the open-source code \textsc{Krome}\footnote{\url{bitbucket.org/tgrassi/krome/commits/372e90a}} \citep{Grassi2014} for the calculations. We compare a run using a full implementation of the dust physics, with each dust-bin evolved dynamically, and a run using the dust tables presented here. In either case the dust temperature and the molecular hydrogen formation is computed on-the-fly consistently with the chemical and thermal evolution of the gas.
 \begin{figure}
  \centering
  \includegraphics[width=.49\textwidth]{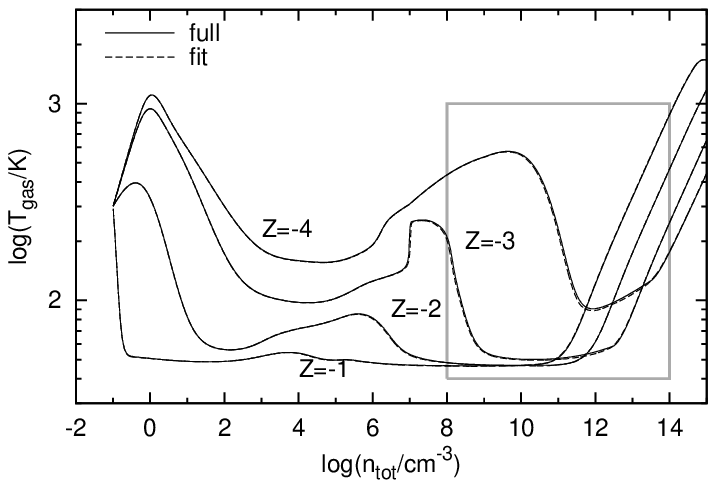} 
  \includegraphics[width=.49\textwidth]{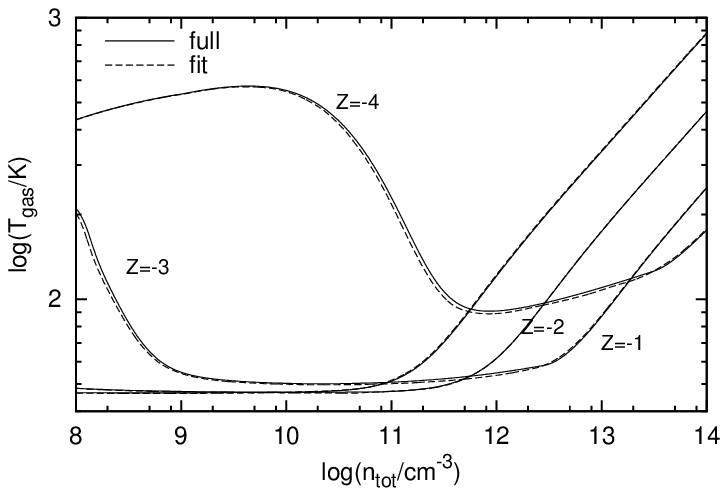} 

  \caption{Application~1. Top panel: comparison between a one-zone collapse test run in \krome using the full dust machinery (solid) and using the tables (dashed) for different metallicities indicated by the labels. Bottom panel: zoom-in on the high-density region contained in the grey box in the upper panel. Notice that the aspect-ratios of the grey box in the top panel and the bottom panel figure are different.}
  \label{fig:comparison}
 \end{figure}

Fig.~\ref{fig:comparison} shows the phase-space evolution of the collapse for different metallicities ($Z=-4$, $-3$, $-2$, and $-1$). Dust cooling is important at higher densities where the temperature quickly drops (the region in the grey box), while the cooling at lower densities is mainly from metals. The H$_2$ formation also affects the thermal evolution for $n_{tot}\gtrsim10^5$~cm$^{-3}$ due to chemical heating from the process \citep{Hollenbach1979}. \tgcomment{As already described in \citet{Omukai2005} and in \citet{Grassi2014}, the four models show  a similar behaviour controlled by the interplay of compressional heating and metal cooling, then, when atomic carbon is converted into CO, the cooling is dominated by H$_2$, while at intermediate densities the temperature increases because of exothermic reactions (including H$_2$ formation on dust). The temperature starts to decrease again when the gas is optically thick and the density is high enough to have gas-dust coupling, but it suddenly stops after the dust evaporation and the compressional heating takes over.}  The results are similar to Fig.~7 in \citet{Omukai2005}, and for this reason we refer the interested reader there for a detailed description of the physics involved in the problem.

Note that the results of the tabulated functions almost overlap with the full dust calculations (error on $T_g$ is $\lesssim1\%$), but the tables are three times faster than the original implementation in \kromes, which is already highly optimised (e.g.~see Appendix~\ref{sect:tdust_dt_full}). Moreover, the tables can handle several types of dust and an arbitrary number of bins without affecting the performance of the numerical simulation, while in the model without the tables the computational cost increases with the complexity of the dust model, resulting in an even a larger speed-up when the problem requires a large number of bins and/or types of dust. This advantage is particularly suitable for the large hydrodynamical simulation such as those discussed in Sects.~\ref{sect:hydro} and~\ref{sect:hydro_cloud} below.

The phase space plot in Fig.~\ref{fig:shaded_Tdust}  indicates the area spanned by the maximum and the minimum dust temperature $T_{d}$ for silicate and carbonaceous dust grains. The difference in temperature between different grain sizes is only present when the metallicity is $Z\leq-3$, as shown by the small and the large shaded area labelled with $Z=-3$ and $Z=-4$, respectively. For $Z\geq-2$ the difference in dust temperature is not visible.
 \begin{figure}
  \centering
  \includegraphics[width=.49\textwidth]{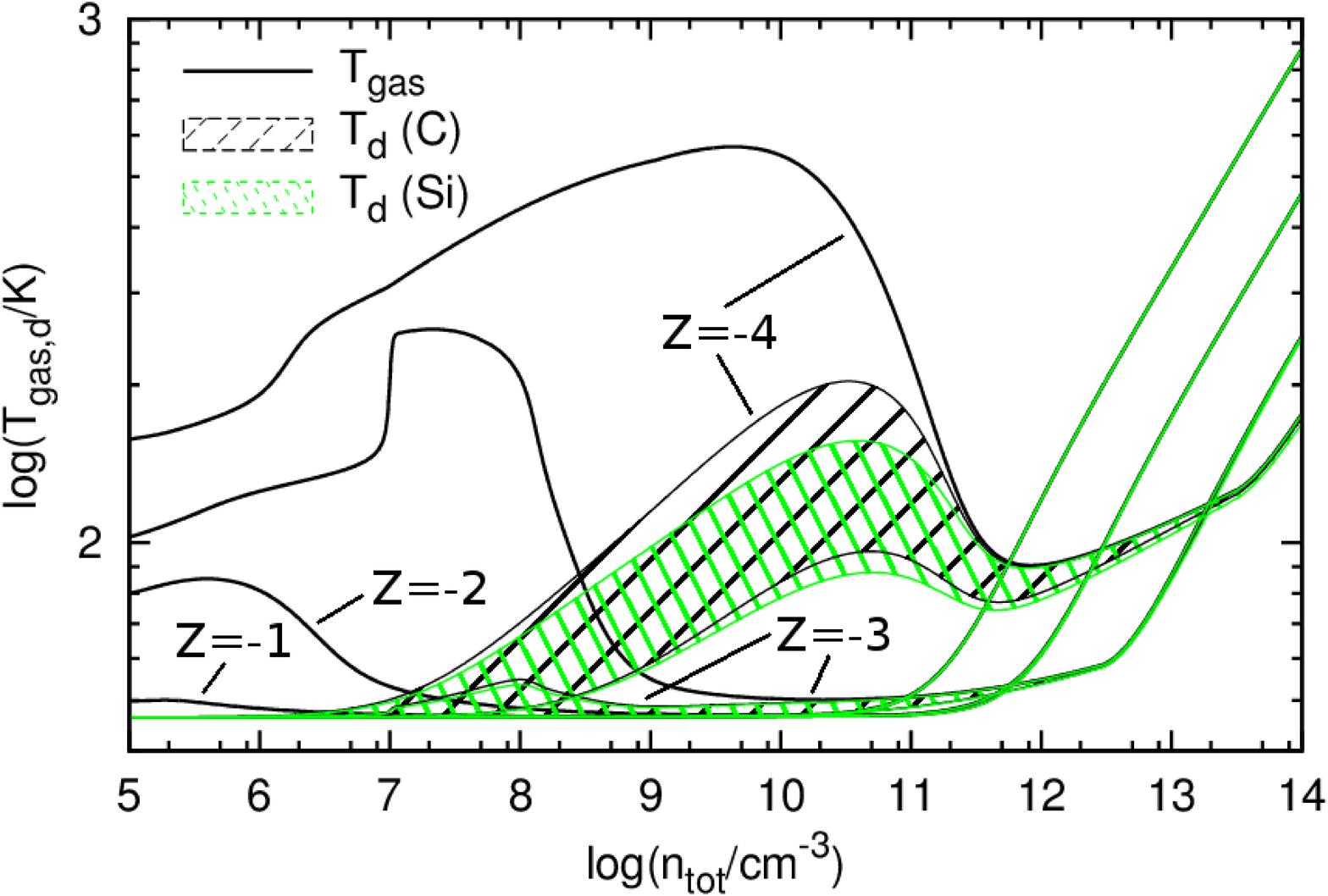} 

  \caption{Application~1. Gas temperature evolution (solid lines) for different metallicities as in the labels. We also include the span in temperature of the different grains both for carbon- (black-hatched) and silicon-based (green-hatched) dust. Note that for $Z=-2$ and $Z=-1$ the effect is not visible, i.e. there is no temperature difference between the dust components.}
  \label{fig:shaded_Tdust}
 \end{figure}
\subsection{Parameter sensitivity}\label{sect:param_sensitivity}
\begin{figure*}
  \centering
  \includegraphics[width=.49\textwidth]{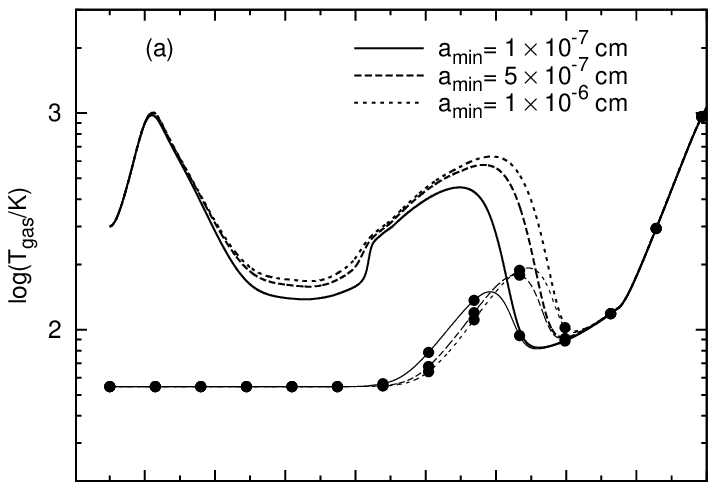} \hspace{-10mm}
  \includegraphics[width=.49\textwidth]{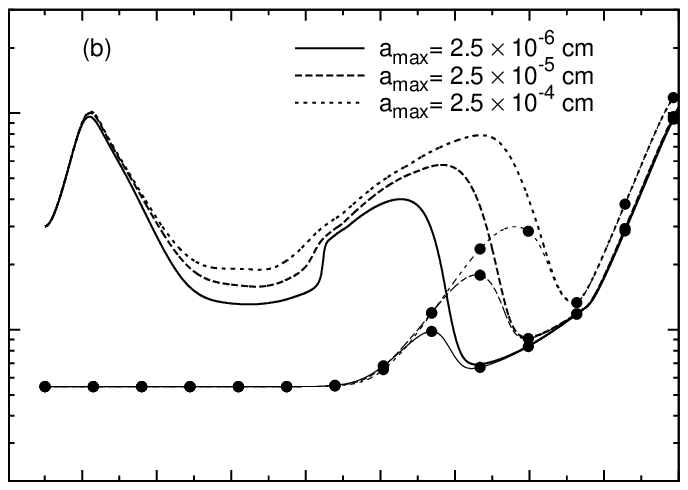}

  \vspace{-5mm}

  \hspace{1mm}\includegraphics[width=.49\textwidth]{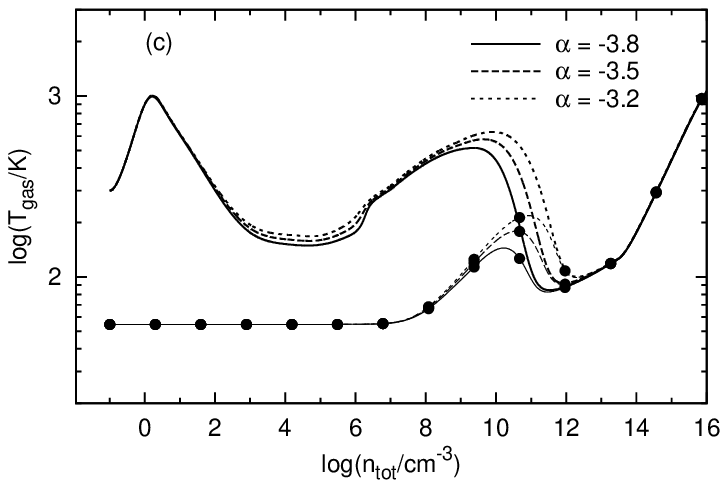}\hspace{-9mm}
  \includegraphics[width=.49\textwidth]{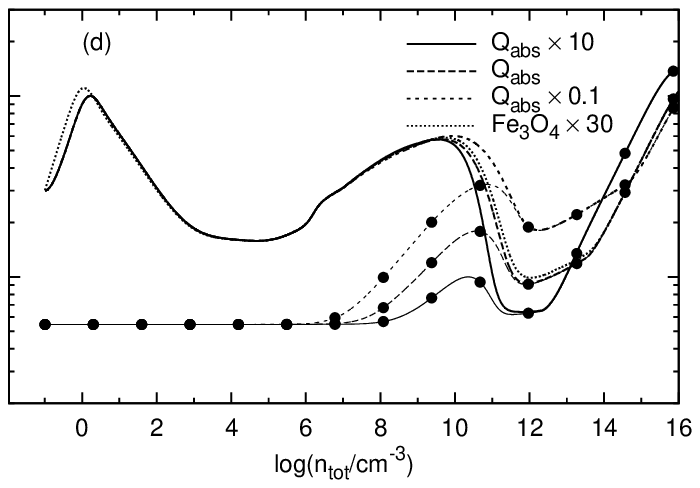}
  \caption{Application~1. Phase-space evolution of the one-zone collapse when some of the key parameters are modified: (a) minimum and (b) maximum size of the dust distribution, (c) power-law exponent, and (d) absorption coefficient. The lines with points are the corresponding dust temperatures for the tenth bin of the carbonaceous grain distribution, assumed a representative bin.}
  \label{fig:parameters}
 \end{figure*}

In this subsection we investigate the same model but varying some of the key parameters, namely $a_{min}$, $a_{max}$, $p$, and $Q_{abs}$, which are increased (or decreased) considerably in order to have an appreciable effect on the thermal evolution. For the tests presented in this Section we use the full dust model, since (in the one-zone context) it is more convenient than producing a complete set of tables for each run.  When the parameters are changed $\rho_d$ is kept constant.
The model employed is the one with metallicity $Z=-4$, but the results are similar for others metallicities. As can be seen in Fig.~\ref{fig:parameters} all the considered parameters affect the gas temperature evolution during the collapse. This effect is mostly due to changes in the total area of the grains in the formation of H$_2$ and in the dust cooling equations. From \eqn{eqn:partial_cooling} and \eqn{eqn:H2formation} we have $\dd n_{\rm H_2}/\dd t\propto a^2n_d$ and $\Lambda \propto a^2n_d$. The total area $A$ for a distribution $\varphi(a)\propto a^p$ is
\begin{equation}
	A(a_{min}, a_{max}, p) = 4\pi\mathcal{C}\int_{a_{min}}^{a_{max}} a^{2+p}\dd a\,.
\end{equation}
Using \eqn{eqn:normalization} to find $\mathcal{C}$ we obtain
\begin{equation}\label{eqn:total_area}
	A(a_{min}, a_{max}, p) \propto \frac{a_{max}^{p+3}-a_{min}^{p+3}}{a_{max}^{p+4}-a_{min}^{p+4}}\,\,\frac{p+4}{p+3}\,,
\end{equation}
where the proportionality term is a positive constant that
depends on the overall density.
The variation of the total area $A$ with the parameters $a_{min}$, $a_{max}$, and $p$ can be analysed by studying the sign of $\partial A/\partial  a_{min}$, $\partial A/\partial  a_{max}$, and $\partial A/\partial  p$, respectively. All these quantities are negative in the intervals considered, implying less cooling when these quantities are increased, in agreement with Fig.~\ref{fig:parameters} panels (a) to (c). Also, $\beta_e$ is affected by the total area, causing a change in the dust temperature found, but this is a
second order effect compared to the area dependence.

The variation of $Q_{abs}$ by one order of magnitude is shown in Fig.~\ref{fig:parameters} (d). In particular, scaling $Q_{abs}$ by a constant value affects the integral of $Q_{abs}(E)J(E)$ in  \eqn{eqn:dust_opacity}, and hence it changes the root $T_{d,i}$ in \eqn{eqn:beta_balance_basic}. When $Q_{abs}$ decreases (i.e.~when the emission efficiency is lower) the coupling with the gas temperature is faster, and the dust temperature increases quicker towards the gas temperature as seen in Fig.~\ref{fig:parameters} panel (d). In the same panel we also plot the thermal evolution including a third dust population, namely Fe$_3$O$_4$ (magnetite). For this grain type we assume a bulk density $\rho_{0}=2.45$~g~cm$^{-3}$ \citep{Turgut1996} and iron as key-element \emph{scaled 30 times} its solar abundance in order to have a visible effect in the plot. The optical properties are calculated as described in Appendix~\ref{appendix:qabs}, with the imaginary refractive index ($n$, $k$) taken from the Jena database by Amaury~\&~Triaud (unpublished\footnote{\url{astro.uni-jena.de/Laboratory/OCDB/mgfeoxides.html\#C}}). Fig.~\ref{fig:parameters} panel (d) shows that the optical properties are dominated by carbon- and silicon-based dust grains and the presence of magnetite has a small effect on the thermal evolution of the gas even when enhanced by a factor of 30.

\begin{figure}
  \centering
  \includegraphics[width=.49\textwidth]{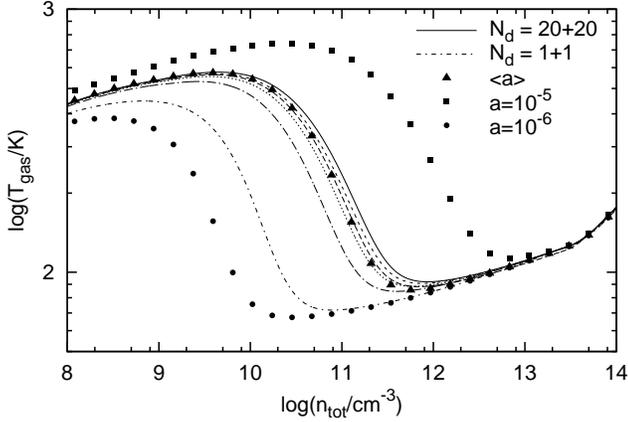} 

  \caption{Application~1. Phase-space evolution of the one-zone collapse using a varying number of dust bins. $N_d=N_{d,{\rm C}}+N_{d,{\rm Si}}$ indicates the number of carbonaceous and silicate dust grain bins used. The lines between 20+20 and 1+1 are for 10+10, 5+5, 3+3, and 2+2 bins respectively. The points represent the evolution calculated with 1+1 bins employing the surface-averaged size of \eqn{eqn:average_size} (triangles), or with an assumed average dust size of $\langle a\rangle=10^{-5}$~cm (squares) or $\langle a\rangle=10^{-6}$~cm (circles).  More details in the text.}
  \label{fig:parameter_bins}
 \end{figure}

The product between the area and the number density plays a key role, hence due to the
less accurate sampling we expect that using a small number of bins will impact the evolution. In Fig.~\ref{fig:parameter_bins} we show how the model converges as a function of logarithmically spaced bins $N_d=N_{d,{\rm C}}+N_{d,{\rm Si}}$ that includes carbonaceous and silicate grains. Convergence is reached for approximately more than $5+5$ bins; however, this value can be different for different environments and gas thermal histories, and can also be improved if a better problem-dependent binning strategy is employed. For a power-law dust distribution, using \eqn{eqn:total_area}, the surface-area averaged dust size is
\begin{equation}\label{eqn:average_size}
	\langle a\rangle = \frac{a_{i+1}^{p+3}-a_{i}^{p+3}}{a_{i+4}^{p+4}-a_{i}^{p+4}}\,\,\frac{p+4}{p+3}\,.
\end{equation}
Using this as the average dust size gives an almost perfect result even with a single bin, as shown in Fig.~\ref{fig:parameter_bins}, labelled $\langle a\rangle$, but this method depends on the process considered (in this case area-depending), and it is still less accurate than the table approach since the latter can employ an arbitrary number of bins and grain types without affecting the computational cost. Therefore we expect that for complicated dust compositions \eqn{eqn:average_size} may not be a good approximation, as discussed in Sect.~\ref{sect:ava} below. Note that \eqn{eqn:average_size} is valid only for power-law dust distributions where $p$ is defined, while our method can be applied to any $\varphi(a)$ function.

Finally, assuming an arbitrary average size (e.g. $\langle a\rangle =10^{-6}$~cm or $\langle a\rangle =10^{-5}$~cm), even if properly normalised to the total dust density, results in a wrong phase-space evolution as shown in Fig.~\ref{fig:parameter_bins}.

\subsection{Caveats with average size}\label{sect:ava}
\begin{figure}
  \centering
  \includegraphics[width=.49\textwidth]{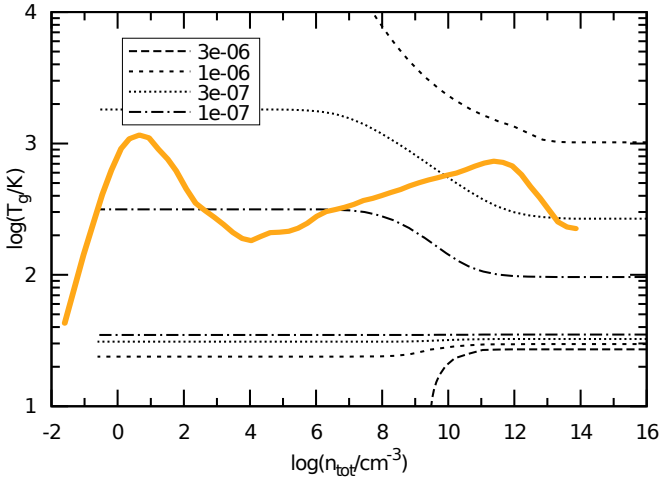} 
  \caption{Contours of the average dust size $\langle a\rangle(T_g,n_{tot})$ found solving \eqn{eqn:avg_system} assuming a silicate grain type (see key, in cm). The values of $f_\Lambda$ and $f_{\mH_2}$ are retrieved from the tables employed in \citet{Bovino2016} for a power-law distribution with $f_{dep}=0.49$ and $p=-3.5$. The thermal profile (orange solid line) is taken from the same work (their Fig.~1, left).}
  \label{fig:contour_ava}
 \end{figure}

In principle, the tables obtained from our calculations can be employed to find an average size dust grain $\langle a\rangle$ and dust temperature $\langle T_d\rangle$ that reproduce both the actual values of cooling and molecular hydrogen catalysis. This is described by the following system of equations
\begin{equation}\label{eqn:avg_system}
\left\{
  \begin{array}{rcl}
    f_\Lambda(T_g,n_{tot}) n_{tot} \mu_g & = & 2\pi(T_g-\langle T_d\rangle) k_B v_g \langle a\rangle^2 n_d   \\
    f_{\mH_2}(T_g,n_{tot}) n_{tot} \mu_g & = & \frac{\pi}{2}  \Phi\left[T_g,\langle T_d\rangle\right] v_g \langle a\rangle^2 n_d\,,
  \end{array}
\right.
\end{equation}
where in each equation the LHS and the RHS represent the table and the averaged quantities, respectively. The terms are defined in \eqn{eqn:rad_balance_bin} and in Sect.~\ref{sect:H2} above, and the unknowns are $\langle a\rangle$ and $\langle T_d\rangle$. The efficiency factor is $\Phi\left[T_g,\langle T_d\rangle\right] = S_{\rm Si}\left[T_g,\langle T_d\rangle\right]\epsilon_{\rm Si}\left[T_g,\langle T_d\rangle\right]$ assuming a silicon grain.
The system can be solved numerically in order to find $\langle T_d\rangle$ and $\langle a\rangle$ for every $(T_g,\,n_{tot})$ pair. With two equations and two unknowns, once the temperature and the density \emph{of the gas} are given, there is only one \emph{unique combination} of $\langle T_d\rangle$ and $\langle a\rangle$ that satisfies \eqn{eqn:avg_system}. In Fig.~\ref{fig:contour_ava} we plot on a phase diagram contour lines of the average size $\langle a\rangle$ as a function of $T_g$ and $n_{tot}$,  found for the mixture of dust discussed in \citet{Bovino2016}, where a $p=-3.5$ power-law distribution of silicon and carbon grains in the range $a=[5\times10^{-7}, 10^{-5}]$~cm, with $f_{dep}=0.49$ and $Z=-4.35$ was used. We note that to mimic the reference model behaviour (both for cooling and H$_2$ formation at the same time), every thermal evolution trajectory that crosses different grains size contour lines (e.g.~solid orange line, taken from Fig.~1, left, in \citealt{Bovino2016}) should use a \emph{non-physical} variable average size, which implies that assuming constant $\langle a\rangle$ will lead to errors in one of the two quantities computed, i.e.~either cooling or H$_2$ surface catalysis. This variation in $\langle a\rangle$ results from solving the set in \eqn{eqn:avg_system} when following the phase diagram track shown in Fig.~\ref{fig:contour_ava}. The size of the error depends on the characteristics of the system and on the assumptions made on the dust properties and composition. By construction, the tables presented in this work are not affected by this problem.

\subsection{Limits of the present methodology}\label{sect:limits}
The tables presented in this work do not depend on the chemical composition of the gas, since they are defined as in \mbox{Eqns.~(\ref{eqn:fH2}-\ref{eqn:fTdust})}. However, all the processes that depend on the history of the gas chemistry, and that require some time-dependent approach, cannot be included in our machinery without simplifications and assumptions. In particular, any process that affects the grain size distribution, as for example growth \citep{Nozawa2012,Chiaki2013}, requires a known \emph{evolution} of the gas abundances and the sticking species (e.g. C for amorphous carbon grains), and the tables (by construction) cannot include this history. Analogously, we cannot track the formation of any ice mantle that would affect the optical properties of the dust \citep{Semenov2003}. Another missing process is the gas-grain interaction in the presence of charged grains, that might change the cooling efficiency \citep{Draine1987} and the chemistry \citep{Semenov2010}. The charge distribution of dust strongly depends on the impinging radiation \citep{Draine1978,Weingartner2001} and on the presence of ions in the gas \citep{Burke1983}. All the processes mentioned can be included though, if assumptions are made about the most probable history a parcel of gas and dust takes to reach a certain point in the $(n_{tot}, T_{gas})$ phase-space, making the method encompass more physics at the cost of reducing the self-consistency.

\section{Application 2: three-dimensional dust-enriched minihalo collapse}\label{sect:hydro}
To assess the feasibility of our approach in 3D hydrodynamical calculations we employ the dust tables to follow the collapse of a minihalo starting from cosmological initial conditions. The details of our numerical setup have been reported in many papers and have recently been discussed in \citet{Bovino2016}, where a similar approach has been used. We allow 27 levels of refinement and high-resolution, resolving the Jeans length by 64~cells. The simulations are performed with the \textsc{Enzo} code \citep{Enzo2013} with the publicly available \krome patch\footnote{\url{bitbucket.org/tgrassi/krome/commits/d74eab9}}. The thermal and chemical evolution of the gas is then solved by \krome instead of using the standard chemistry provided in \textsc{Enzo}. We follow the evolution of nine chemical species: H, H$^+$, H$_2$, H$_2^+$, H$^-$, He, He$^+$, He$^{++}$, and electrons.  The system of differential equations includes 27~kinetic reactions using the most recent compilation of rate coefficients, among others the new three-body H$_2$ formation by \citet{Forrey2013}. Thermal processes include atomic cooling and H$_2$ roto-vibrational cooling, which has been updated to the new data reported by \citet{Glover2015}. The dust tables have been prepared to obtain the highest accuracy and convergence in terms of dust bins: we have included 20 carbon 
 (graphite) and 20 silicon (astronomical silicates olivine-like) bins. A standard MRN power law ($p= -3.5$) distribution has been used with grain sizes between 5$\times$10$^{-7}$ and 10$^{-5}$ cm. We use a fixed uniform metallicity of $Z = -4$  and a dust-to-gas ratio $D = 0.00934\,10^Z$, i.e. a linear re-scaling of the Milky Way dust abundance\footnote{See \citet{Klessen2014} for a discussion about the caveats related to this assumption.}. It is important to note that no additional modifications are required to \textsc{Enzo} as we do not have any additional passive scalars. Once the thermal evolution has been obtained, the dust temperature can be evaluated \emph{a posteriori} as described in Sect.\ref{sect:physics_bin_notation}.
A redshift of $z=16$ is assumed to reduce the dimensionality of the tables, since the CMB radiation is redshift-dependent, as indicated by $J_z$ in \eqn{eqn:rad_balance}. The error caused by this assumption is negligible as most of the evolution of the minihalo occurs around $z = 16$, after having reached the virial stage.

 \begin{figure}
  \centering
  \includegraphics[width=.5\textwidth]{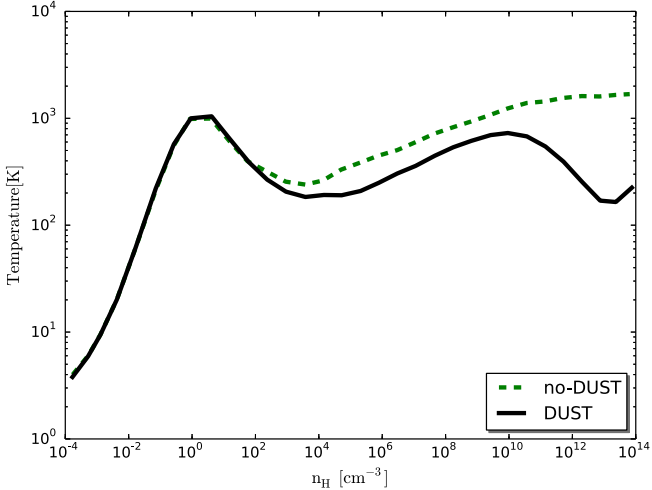} 
  \caption{Application~2. Average temperature evolution as a function of the total hydrogen nuclei number density for the run with dust (solid black) and without (dashed green). See text for more details.}
  \label{fig:enzo_phase}
 \end{figure}

 \begin{figure}
  \centering
  \includegraphics[width=.5\textwidth]{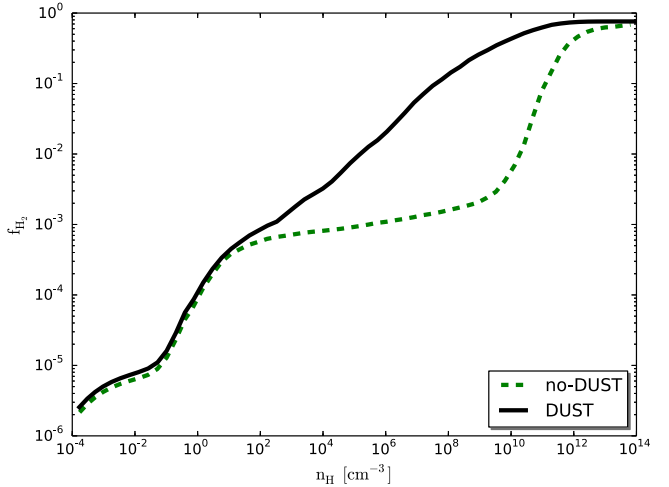} 
  \caption{Application~2. Averaged mass fraction of molecular hydrogen as a function of the total hydrogen nuclei number density for the model with dust (solid) and without (dashed). }
  \label{fig:enzo_H2}
 \end{figure}

 \begin{figure}
  \centering
  \includegraphics[width=.5\textwidth]{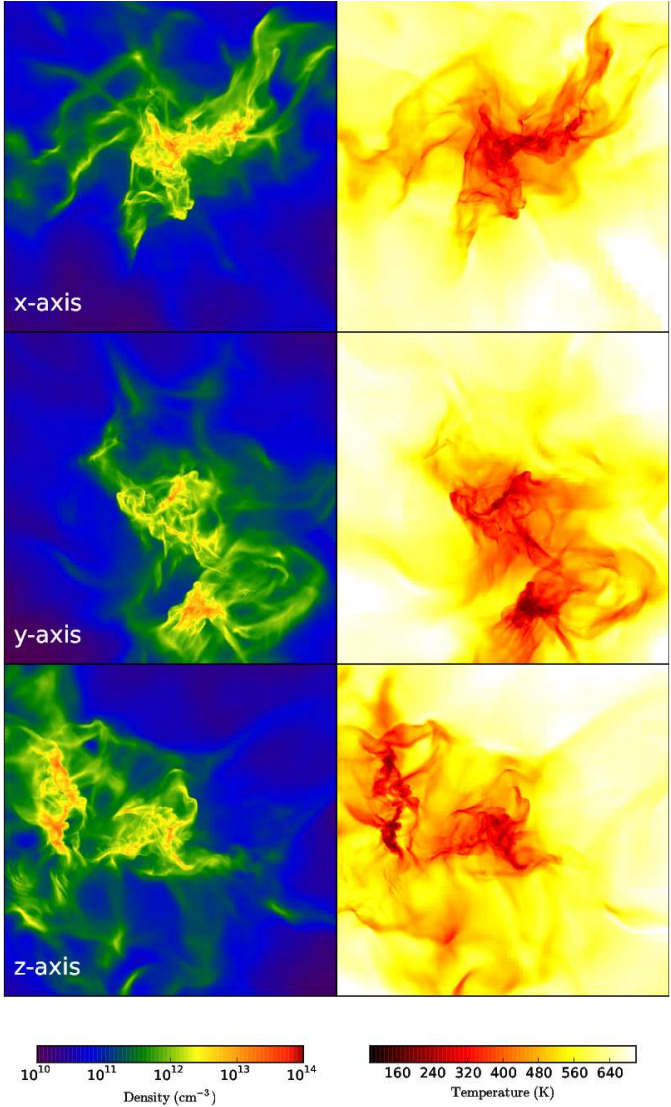} 
  \caption{Application~2. Density (left) and temperature (right) projections along x, y, and z axes, for the run including dust. Each panel spans $200$~AU. }
  \label{fig:enzo_projections}
 \end{figure}

We report in Fig.~\ref{fig:enzo_phase} the average temperature as a function of the total hydrogen nuclei number density $n_{\rm H}$ obtained from the collapse of a dust-enriched minihalo. The evolution  resembles the results shown in Sect.~\ref{sect:onezone} for the one-zone model. In particular, we can distinguish three different features: (i) the halo virializes around 1~cm$^{-3}$ and then starts to cool due to H$_2$. (ii) At densities larger than $10^5$~cm$^{-3}$ the formation of H$_2$ on dust becomes efficient and produces a chemical heating that slightly rises the gas temperature. (iii) At higher densities coupling between gas and dust grains starts to be effective and the dust cooling kicks in bringing the temperature down to $\sim150$~K. We also observe the transition between the optically thin and the optically thick regime around a density of $10^{13}$~cm$^{-3}$ as an effect of the dust opacity. At this stage gas and dust temperatures are fully coupled. In the same Figure we also compare with the average temperature obtained from a run without dust (see e.g. \citealt{Bovino2013a}). As for the one-zone collapse discussed in Sect.~\ref{sect:onezone} the difference between a pure primordial case ($Z=-\infty$) and the run with $Z=-4$ is remarkable. Formation of H$_2$ on dust already has an effect at intermediate densities ($10^3 < n < 10^6$~cm$^{-3}$) and the dust cooling becomes very efficient at densities higher than $10^{10}$~cm$^{-3}$, where the interaction between gas and dust grains increases. In Fig.~\ref{fig:enzo_H2} we report the H$_2$ mass fraction as a function of $n_{\rm H}$. H$_2$ forms much more rapidly in the presence of dust as already reported in previous works (e.g. \citealt{Omukai2000,Omukai2005,Smith2015}).

Finally, Fig.~\ref{fig:enzo_projections} shows density and temperature projections along the three axes (x, y, and z) for the run where dust was included taken at the maximum peak density (\mbox{$\sim$10$^{-10}$ g cm$^{-3}$}). The dust induces a strong filamentary structure as recently reported by \citet{Bovino2016}. The temperatures in the central region are around 150~K as also seen in Fig.~\ref{fig:enzo_phase}.

Overall the table-based approach to describe the impact of dust on the gas chemistry and thermodynamics included in these high-resolution simulations increased the computational time by $\sim20$\% compared to a standard primordial (no-dust) run. We have to consider that the presence of the dust affects the physical conditions, so that a real comparison is not possible, and we refer the reader to the speed-up obtained in Sect.~\ref{sect:onezone}, where a proper methods comparison has been made. However, in the context of the 3D hydrodynamical simulations this is a remarkable result considering the detailed dust physics (distribution, grain types, number of grain bins, optical properties) employed. In addition, as already stated in previous sections, this approach allow us to add an arbitrary number of bins and dust types without affecting the final computational cost.

\section{Application 3: Dust cooling and H$_2$ formation in solar metallicity molecular clouds}\label{sect:hydro_cloud}
The structure and dynamics of molecular clouds is determined to a large extent by supersonic turbulence, and in particular
the strength and form of the turbulence compared to the potential energy plays a key role for the star formation rate, and the
initial mass function \citep{Padoan2002,PadoanPPVI,Padoan2014,Federrath2012}. MHD simulations including either stellar feedback \citep{Padoan2016} or
artificial large-scale driving \citep{Padoan2014} can be used to study the development of molecular clouds. Traditionally, in such models the gas is often
assumed to be isothermal, but a precise account of the thermodynamics is needed for models that contain patches of warmer gas, or follow the
gravitational collapse of individual cores down to the formation of individual protostars. Furthermore, long wavelength observatories have in recent
years given us rich data sets at many different scales from parsecs to AUs. A consistent microphysical model is important
to trace the dust temperature and chemical abundances to compare with observations and determine the temperature of the cloud environment through
a precise account of relevant heating and cooling processes. Within this context in this section we present a new simulation of an outer Galaxy like patch
of a molecular cloud at high numerical resolution including
a comprehensive chemical model for reactions among H-C-O molecules.  The chemical model is similar to \citet{Glover2010},
but has been updated with the newest rates from e.g.~the KIDA database, and it is contained
in the file \verb+react_COthin_noSi+ in \kromes\footnote{\url{bitbucket.org/tgrassi/krome/commits/c09e1bd}}, where references for
the different rates can be found. We consider 277 reactions and the following 34 species:
H, H$^+$, He, He$^+$, He$^{++}$, H$_2$, H$_2^+$, H$^-$, C$^+$, C, O$^+$, O,
OH, HOC$^+$, HCO$^+$, CO, CH, CH$_2$, C$_2$, HCO, H$_2$O, O$_2$, H$_3^+$, CH$^+$, CH$_2^+$, CO$^+$,
CH$_3^+$, OH$^+$, H$_2$O$^+$, H$_3$O$^+$, O$_2^+$, C$^-$, O$^-$, and electrons. The gas cooling is provided
by H$_2$, CO, CI, CII, and OI lines, endothermic reactions, and continuum, while heating processes includes cosmic rays,
photoheating ($A_v$ based), exothermic reactions, and photoelectric heating from dust grains scaled with $A_v$, the last
as in \citet{Seifried2016}. All these processes are solved self-consistently. For the details of the physical processes
we refer the reader to \citet{Grassi2014} and the references therein.

To illustrate the importance of including a self-consistent calculation of dust effects we compare below a model where H$_2$ formation
on grains is described as a simple rate equation following \citet{Hollenbach1979,Glover2010} with a model where instead the table-based
approach described in the previous sections is employed to calculate dust cooling and formation of H$_2$ with
the approximation of $A_v$ defined in Sect.~\ref{subsub:ism}, assuming $\alpha=2/3$. We have made this empirical relation by fitting a power-law above number densities of
200~cm$^{-3}$ to the density-$A_v$ relation from the hydro-chemical models including full radiative transfer by \citet{Glover2010}. Recently, a similar approach has been discussed by \citet{Safranek2016}.
Thanks to this approximation the tables are functions of $n_{tot}$ and $T_{gas}$ only. However, they can be extended to consider a generic $A_v$ by using
a trilinear interpolator. These extended tables would then be suitable for models that compute $A_v$ directly with
ray-tracing techniques or equivalent methods.

The models are computed using a modified version of the \textsc{Ramses} code \citep{Teyssier2002}. See e.g.~\citet{Padoan2014,Frimann2016} for
recent descriptions of the modifications. The patch for interfacing the chemistry to \textsc{Ramses} is distributed with {\kromes}. In addition, we have
made extensive changes
to support a variable adiabatic index, both when converting between pressure and internal energy, and when calculating states in the Godunov solvers.
To ensure the conservation of atomic species when refining cells and advecting passive scalars we use a consistent interpolation method described in
Appendix~\ref{appendix:advection}. The MHD variables are evolved using a second order MUSCL scheme with a high precision HLLD solver
and mon-cen slope limiter. Interfaces towards cells where the fast-mode speed exceeds 65~km~s$^{-1}$, are solved using a much more diffusive
local Lax-Friedrich solver. This is important to avoid that a few exceptional cells in the model severely limit the timestep. It typically occurs in low-density
regions with strong magnetic fields, and is only active in $\sim$0.01\% of the cells.

To model conditions reminiscent of present day molecular clouds, as observed in local star forming regions, we use a (3.3~pc)$^3$ box with
an average density of $\rho_0=4\times10^{-21}$~g~cm$^{-3}$ corresponding to a equivalent molecular hydrogen number density of 1200~cm$^{-3}$, and an initially
homogeneous magnetic field of 7~$\mu$G. We use a cosmic ray flux of $\zeta_{\textrm{CR}}=1.3\times10^{-17}$~s$^{-1}$. The strength of the UV interstellar
radiation field is set to 1.69 in units of the Habing flux. The metallicity is solar with
an initial mass abundance set to $X_{\rm H}=0.75236$, $X_{\rm H^+}=8.2\,10^{-5}$, $X_{\rm He}=0.24253$, $X_{\rm H_2}=1.505\,10^{-6}$, $X_{\rm C}=0.001262$,
and $X_{\rm O}=0.003773$, while all other species have zero initial abundance, except electrons that are set to ensure charge neutrality. The temperature
is set to $T / \mu = 20$~K. Rapidly, this initial state evolves towards a balance between heating from UV, cosmic rays and compression due the driven turbulence,
and  cooling processes, resulting in a medium dominated by molecular hydrogen and neutral helium, and a mass-weighted average temperature of
$\sim10$~K. Following \citet{Padoan2014}  we drive the turbulence randomly on the largest scales using a solenoidal acceleration reaching at saturation
an rms velocity of $\sim2$~km~s$^{-1}$ corresponding to a sonic Mach number $\mathcal{M}_s=11$ at 10~K, which is in good agreement with the Larson
size-velocity relation for local molecular clouds \citep{Heyer2009}.

To realise a fully turbulent medium with no memory of the initial conditions, we first drive for 6 turn-over times
using a root-grid resolution of 128$^3$ computational cells, where $t_\textrm{turn-over}= L_\textrm{box} / (2 v_\textrm{RMS}) = 0.79$~Myr.
Afterwards, to better resolve the turbulence and chemical evolution in the densest molecular cloud fragments, we add 6 AMR levels and apply refinement
criteria based on both density and gradients in density, pressure, velocity, and magnetic energy reaching a maximum resolution equivalent to $8192^3$ or
800~AU. The model runs with refinement enabled for one full turn-over time, which is enough to develop the turbulence in the refined cells and reach a well-mixed
turbulent and chemical state at the higher resolution. The total run-time is 5.6 Myr. We use a
``Truelove criterion'' \citep{Truelove1997} and refine one level for each factor of 4 increase in density. On the first three AMR levels we in addition refine if relative gradients
of 2.9, 3.8, 4.0, or 2.4 are detected in either density, pressure, velocity, or magnetic energy. With these refinement conditions there is a total of $\sim$15 million
cells in the model, with the bulk of them located at the three first AMR levels above the root grid.

\begin{figure}
  \includegraphics[width=.47\textwidth]{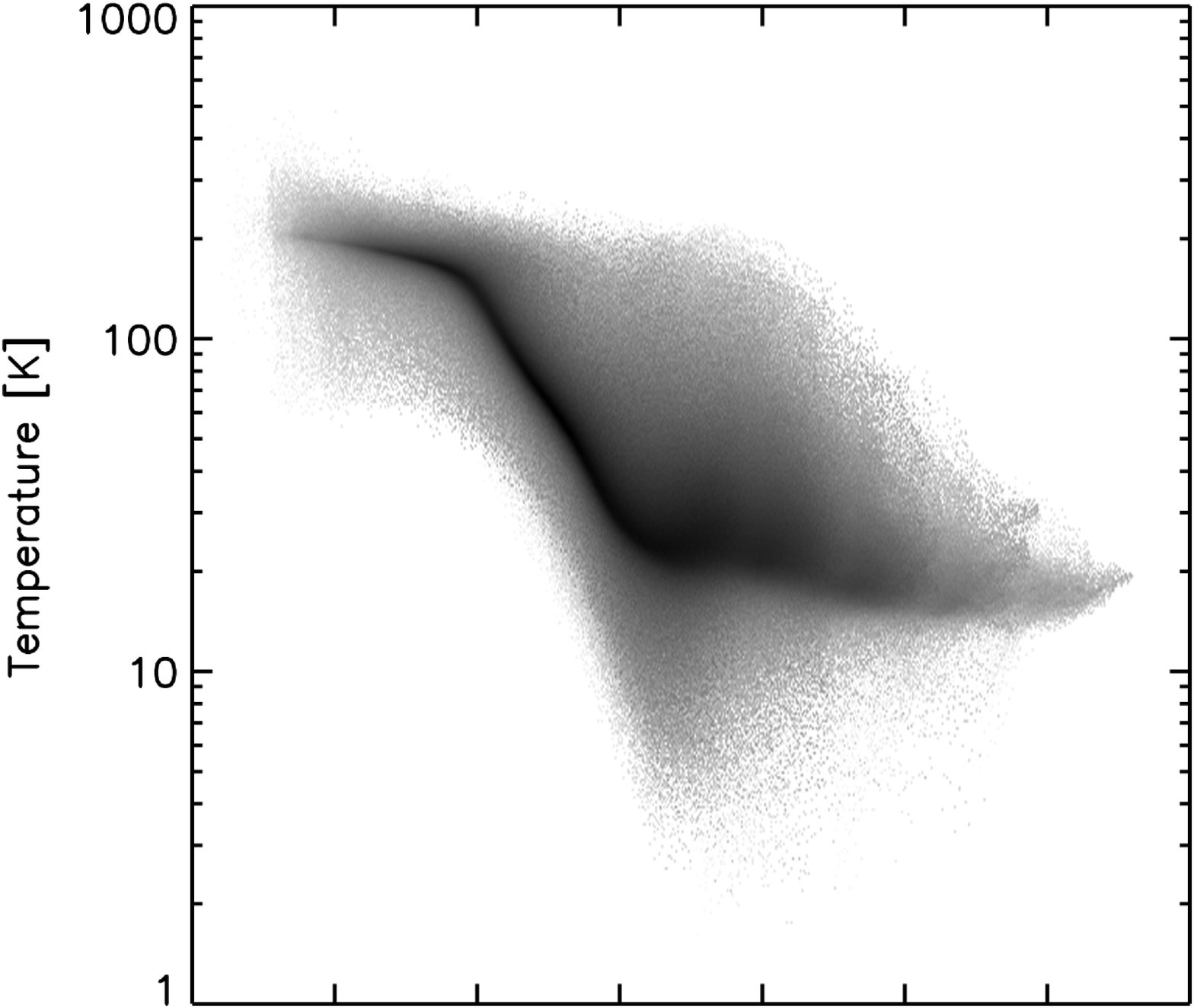} 
  \includegraphics[width=.47\textwidth]{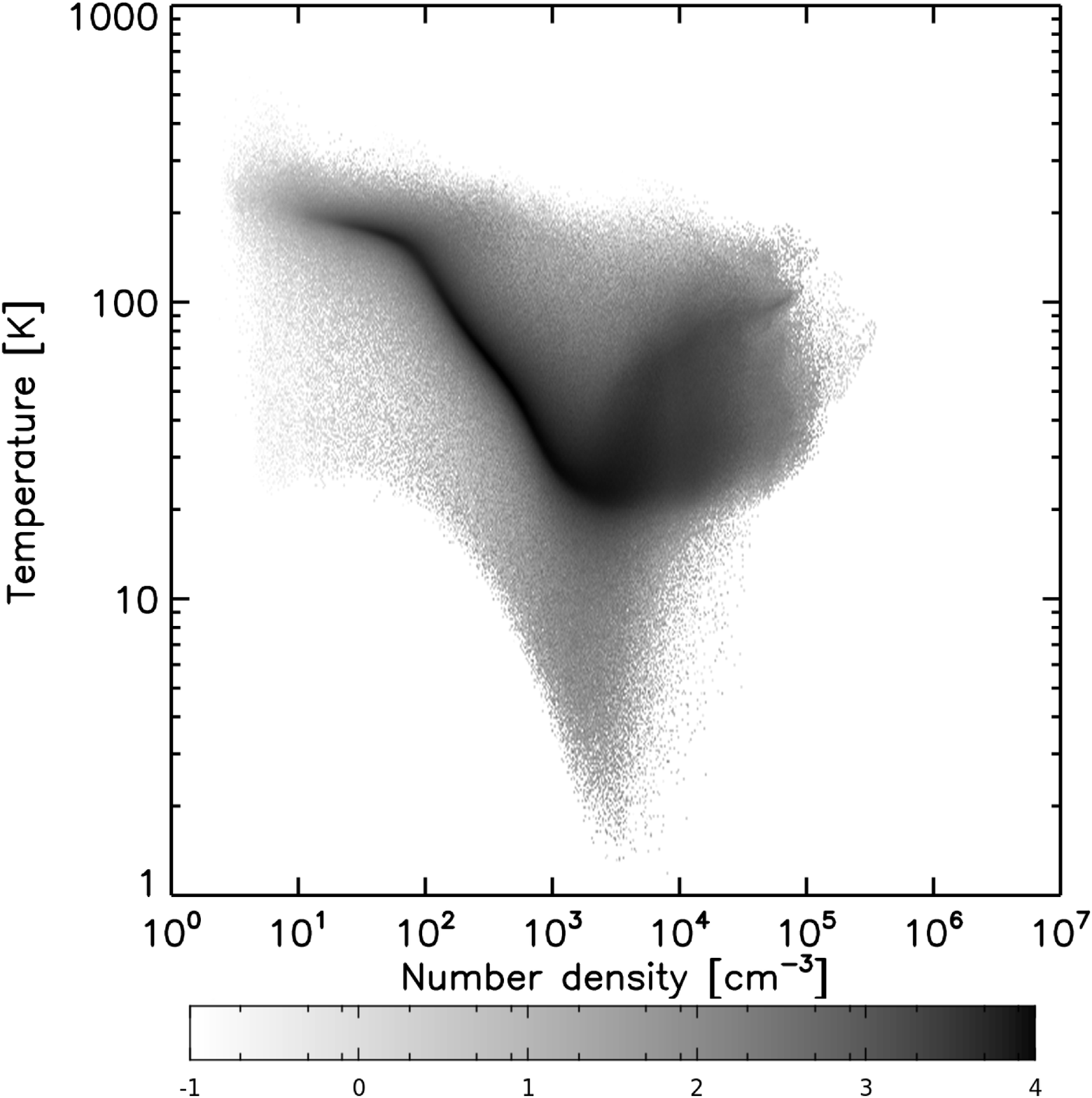}
  \caption{Application~3. Mass weighted number density-temperature phase-space density for models with (top) and without (bottom) the table-based description of dust cooling and H$_2$ grain formation. The color scale is logarithmic in the phase-space density in units of M$_{\odot}$~dex$^{-2}$. Each pixel is $10^{-4}$~dex$^2$. See text for more details.}
  \label{fig:ramses_phase}
 \end{figure}

Shown in Fig.~\ref{fig:ramses_phase} are phase-space distributions with (top) and without (bottom) the table-based approach. The behaviour is analogous to the other tests shown in the previous sections (see e.g. Fig.~\ref{fig:comparison}), where after an initial cooling at low densities (in this case from H$_2$ and metals) for both the models, the main difference comes from the inclusion of
dust cooling, which completely changes the thermodynamics of the high density gas above $10^4$~cm$^{-3}$, by reducing the temperature by $\sim80$~K, in good agreement with one-zone collapse models. The heating from molecular hydrogen formation is barely recognisable in the top panel around $10^4$~cm$^{-3}$, where a small temperature increase is present. In parts of the model divergent flows drives the temperature dynamically below the CMB temperature. This is only visible in the phase-space diagram because of the logarithmic color-scale and happens at exceptional places. Only 0.05\% of the volume is at temperatures below $T_{\rm CMB}$.

\begin{figure}
  \includegraphics[width=.47\textwidth]{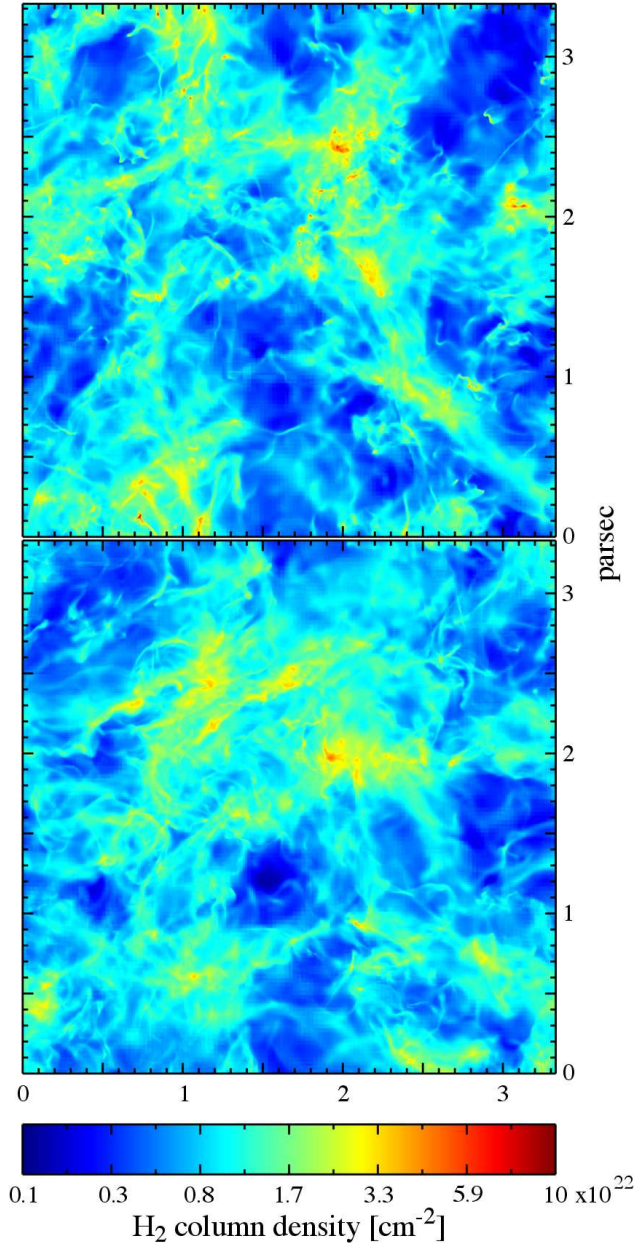}
  \caption{Application~3. H$_2$ column density for models with (top) and without (bottom) the table-based description of dust cooling and H$_2$ grain formation. The color scale is stretched with an exponent of 0.2 to bring out details in lower column depth gas.}
  \label{fig:ramses_col}
 \end{figure}

\begin{figure}
  \includegraphics[width=.47\textwidth]{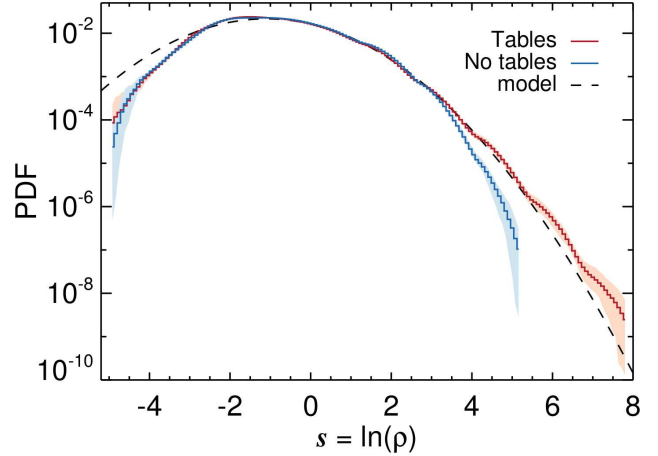}
  \caption{Application~3. PDF of the normalised logarithmic density distribution for the two runs. The dashed line is a model for the PDF at high densties based on \citet{Nolan2015}.}
  \label{fig:ramses_pdf}
 \end{figure}

In Fig.~\ref{fig:ramses_col} is shown the H$_2$ column densities. When dust cooling is included, the lower temperatures at high densities result in more fragmentation with the high density gas residing in thinner filaments and denser cores, solely driven by the turbulence. To quantify the difference in density
structure, in Fig.~\ref{fig:ramses_pdf} is shown the PDF of the mass density in the two cases. For a supersonic turbulent gas
the density distribution is known to be approximately log-normal, and we therefore plot
\begin{equation}
s = \ln( \rho / \rho_0)\,.
\end{equation}
The shaded areas are obtained by calculating $s$ for 11 snapshots selected during the last half turn-over time of the experiment and encloses the minimum
and maximum values for each bin. We selected this time-period because it is after running with AMR turned on for half a turn-over time, and the turbulence and
chemistry of the run should be close to relaxed in the refined cells, while still giving
a time-span to sample the variance. Nonetheless, to robustly measure the variance several turn-over times would be needed, and the shaded areas
almost certainly underestimate the intermittency of the PDF.

The dashed line shows a log-normal model for the PDF
\begin{equation}
\textrm{PDF}(s) = \frac{1}{\sqrt{2\pi\sigma_s^2}}\exp\left(-\frac{(s + \sigma_s^2 / 2)^2}{2 \sigma_s^2}\right)\,.
\end{equation}
For an isothermal gas the variance $\sigma_s$ can be found as \citep{Padoan2011,Federrath2012}
\begin{equation}
\sigma_s^2 = \ln\left(1 + b^2 \mathcal{M}_s^2 \frac{\beta}{1 + \beta}\right) = \ln(1 + \alpha_\textrm{turb}) \,,
\end{equation}
where $\beta= P_\textrm{th} / P_\textrm{mag}=0.53$ is the initial plasma beta, $b=1/3$ depends on the type of the driving, in this case
solenoidal. Examining the phase-space diagram (Fig. \ref{fig:ramses_phase}) suggests that in the
model where dust cooling is included the gas at high densities is slightly soft, with a negative slope giving an effective adiabatic
index of $\gamma=0.9$. Measuring the volume weighted sonic Mach number in the corresponding high density cells with a density
$\rho > 10\, \rho_0$ we find $\mathcal{M}_s(\rho> 10\, \rho_0)=12.4$. \citet{Nolan2015} recently derived a general expression for the variance
in a non-isothermal gas
\begin{equation}
\sigma_s^2 = \ln\left(1 + \frac{(\gamma + 1)\alpha_\textrm{turb}}{(\gamma - 1)\alpha_\textrm{turb} + 2}\right) \,,
\end{equation}
which in our case with the above values yields $\sigma_s^2 = 2.2$, giving an excellent fit to the PDF at high densities, as shown in
Fig.~\ref{fig:ramses_pdf}. At low densities the PDF is shallower. This is easily understood as a consequence of the higher temperature
and stiff behaviour of the gas at those temperatures.
At high densities the PDF for the model without dust tables is significantly shallower, which is consistent with the differences seen in the
phase-space distribution and morphology of the column density shown in Figs.~\ref{fig:ramses_phase} and \ref{fig:ramses_col}.

Including a proper treatment of dust cooling in a hydro-chemical model of the ISM is clearly important not only for the thermodynamics, but also the
structure of the molecular cloud in the model and therefore the stellar population. Models with a self-consistent description of dust, including the cooling,
will have a core mass function that peaks at lower masses and reaches higher densities, resulting in a larger population of low mass stars if self-gravity
and sink particles are enabled in the models.

\section{Conclusions}\label{sect:conclusions}

We have presented a novel methodology based on tables that allows the inclusion of consistent and detailed dust modelling into state-of-the-art hydrodynamical simulations at a negligible computational cost. These tables provide dust cooling, H$_2$ formation rates on grain surface, and dust temperature, as functions of the thermal properties of the gas (temperature and density) and the radiation field. They are based on a self-consistent physical model, and can be extended to an arbitrary number of dust bins and dust types without any additional computational cost for the framework model.
The model is limited in that it assumes a fixed dust distribution, though in principle this distribution could also be a function of the radiation field, and the density and temperature of the gas, as illustrated by our inclusion of dust evaporation. Furthermore, we do not yet include the possible charging of the grains that could significantly change the gas-grain chemistry, and we limit our gas-grain chemistry to that of H$_2$ formation. In the future we will explore the feasibility of including more gas-grain chemistry, such as molecule formation and freeze-out, in our methodology.

We have investigated the consistency of the method using a one-zone model of a gravitational collapse (Application~1). We tested the efficiency of the machinery, and illustrated the importance of proper dust physics for the thermodynamics of the past and present ISM, by using the \textsc{Enzo} code to study a 3D gravitational collapse (Application~2) and \textsc{RAMSES} to model an evolving molecular cloud (Application~3). All tests employ the publicly available code \krome to embed state-of-the-art microphysics and chemistry solved self-consistently and on-the-fly without loosing accuracy.
Although we employed \krome in all our calculations, the methodology presented here is independent from \krome and can be easily incorporated in to any numerical model by using a generic linear interpolator.

Additionally, we have explored the effect of changing several parameters in the one-zone collapse test, in particular analysing different size distribution ranges, slopes, and binning, as well as testing the effect of different optical properties. These tests also showed that if an average grain size is assumed, it is not possible to simultaneously produce accurate and consistent cooling rates and molecular hydrogen formation efficiencies, unless the average grain size has an unphysical dependence on the temperature and density of the gas, which would effectively be a convoluted reconstruction of our tables.

\tgcomment{As shown in the tests presented here, dust plays a key role in many astrophysical environments studied with numerically expensive simulations: dust cooling and/or molecular hydrogen formation on grain surface is important both for the thermal evolution and dynamics, but models often employ a simplified (or no) grain physics that affecting the results. For this reason, having a consistent and fast dust model, as the one presented in this work, is important to properly account for the main effects of the dust. Potential applications that requires intensive 3D models are black hole formation in massive primordial halos polluted by metals and dust \citep{Latif2015}, physics of filaments in star-forming regions \citep{Banerjee2006,Federrath2016,Seifried2016}, star-formation in galaxies \citep{Gnedin2011}, and galaxy mergers \citep{Capelo2015,Mayer2015}. }

Finally, we note that, as an extension of the mini-halo collapse test (i.e.~Application~2), the methodology to create the tables presented in this work has already been employed in \citet{Bovino2016}, where the fragmentation efficiency in a low-metallicity star-forming environment has been studied with a 3D model, assuming standard power-law and more complicated size distributions generated by primordial SNe. As shown in the present work, \citet{Bovino2016} reported that the computational cost of the tables is the same for any arbitrary dust composition and distribution, which makes this methodology particularly suitable for 3D hydrodynamical simulations.

\section*{Acknowledgement}
This research was supported by a Sapere Aude Starting Grant from the Danish Council for Independent Research to TH,
Research at Centre for Star and Planet Formation is funded by the Danish National Research Foundation (DNRF97).
We acknowledge PRACE for awarding us access to the computing resource CURIE based in France at CEA  that was used to carry out
the \textsc{RAMSES} simulations.
SB thanks for funding through the DFG priority program ``The Physics of the Interstellar Medium'' (projects BO 4113/1-2). Figures~6, 7, and 8 of this paper have been obtained by using the YT tool \citep{Turk2011}.
DRGS thanks for funding through Fondecyt regular (project code 1161247), Basal PFB-06 CATA, and through the ``Concurso Proyectos Internacionales de Investigaci\'on, Convocatoria 2015'' (project code PII20150171).
SB and TH acknowledge the kind hospitality of the Kavli Institute for Theoretical Physics (KITP) where this work has been completed. This research was supported in part by the National Science Foundation under Grant No. NSF PHY11-25915.
\bibliographystyle{mn2e}
\bibliography{mybib}

\appendix
\section{Opacity tables}\label{appendix:qabs}
The tables of the absorption coefficient $Q_{abs}$ for carbonaceous and silicates have been taken from Bruce Draine's website\footnote{\url{astro.princeton.edu/~draine/dust/dust.diel.html}}, while for grain types other than these we employed the imaginary refractive indexes from the Jena Database\footnote{\url{astro.uni-jena.de/Laboratory/OCDB/}}. For a sphere of radius $a\ll \lambda = ch/E$ the refractive index is related to the absorption coefficient via the following equation from \cite{DraineBook2011} (Sects.~22.1 to 22.3)
\begin{equation}
	Q_{abs}(a,E) = \frac{C_{abs}(a,E)}{\pi a^2} = \frac{24\pi\epsilon_2}{(\epsilon_1+2)^2+\epsilon_2^2}\frac{a E}{hc}\,,
\end{equation}
where $E=ch/\lambda$ is the photon energy, while the dielectric function is related to the imaginary refractive index as $\epsilon_1+i\epsilon_2 = (n+ik)^2$, leading to   $\epsilon_1 = (n^2-k^2)$ and  $\epsilon_2 = 2nk$.

For grains of size $a\gtrsim \lambda = ch/E$, where the electric dipole approximation is no longer valid, we use Mie theory (Sect.22.5 of \cite{DraineBook2011}) available in the \textsc{scatterlib} package\footnote{\url{code.google.com/p/scatterlib/wiki/Spheres}} and based on \cite{BohrenBook1983} Appendix~A (see also \citealt{Budaj2015}).

\section{Molecular hydrogen formation efficiency on dust grains}\label{appendix:epsilon} 
The efficiency for carbonaceous grain \citep{Cazaux2009} is
\begin{equation}
 \epsilon_{\rm C}\left[{T_g,T_d(a)}\right]=\frac{1-T_H}{1+0.25\left(1+\sqrt{\frac{E_c-E_s}{E_p-E_s}}\right)^2}\exp\left[\frac{E_s}{T_d(a)}\right]\,,
\end{equation}
where $T_H$ is a function of $T_g$
\begin{equation}
 T_{H}=4\left(1+\sqrt{\frac{E_c-E_s}{E_p-E_s}}\right)^{-2} \exp\left(-\frac{E_p-E_s}{E_p+T_g}\right)\,,
\end{equation}
where $E_p=800$ K, $E_c=7000$ K, and $E_s=200$ K.
Analogously, for silicates we have
\begin{equation}
 \epsilon_\mathrm{Si} = \left[1+16\frac{T_d}{E_c-E_s}\,\exp\left(-\frac{E_p}{T_d}-\beta a_{pc}\sqrt{E_p-E_s}\right)\right]^{-1}+ \mathcal{F}\,,
\end{equation}
where $E_p=700$ K, $E_c=1.5\times10^4$ K, $E_s=-1000$ K, $\beta=4\times10^9$,  $a_{pc} = 1.7\times10^{-10}$ m (Cazaux, priv.~comm.), and
\begin{equation}
 \mathcal{F} = 2\frac{\exp\left(-\frac{E_p-E_s}{E_p+T}\right)}{\left(1+\sqrt{\frac{E_c-E_s}{E_p-E_s}}\right)^2}\,.
\end{equation}

\section{Dust temperature differential} \label{sect:tdust_dt_full}
\krome supports two modes of modelling the impact of dust. Either by using a table-based approach, as detailed in this paper, or by tracking dust dynamically as part of the ODE system. In the second case, to increase the stability of the solver and to reduce its internal time-step, \krome tracks the evolution of the dust temperatures as additional independent variables in the system of differential equations (i.e. alongside the species and the gas temperature). These differentials can be found by applying the operator $\dd /\dd t$ on both sides of \eqn{eqn:beta_balance}, which for the \ith grain is
\begin{eqnarray}
	\beta_e\frac{\dd}{\dd t}\int B(E,T_d)Q(E,a) \dd E &=& \beta_e\frac{\dd}{\dd t} \int J(E)Q(E,a) \dd E\nonumber\\ 
		&+& A\frac{\dd}{\dd t}(T_g-T_{d,i})\,,
\end{eqnarray}
where the first terms of the RHS is zero, being constant within the hydrodynamical time-step (i.e. the time interval where \krome is applied). Note that the term $\beta_e(T_d)$ depends on the dust grain temperature, but it is dominated by the temperature of the other bins, so for this reason we can assume that its variation with time (within a solver time-step) is negligible. We apply the chain rule to the LHS in order to obtain
\begin{equation}
	\beta_e\frac{\dd T_{d,i}}{\dd t}\int \frac{\dd B(E,T_{d,i})}{\dd T_{d,i}}Q(E,a) \dd E = A\left(\frac{\dd T_g}{\dd t}-\frac{T_{d,i}}{\dd t}\right)\,,
\end{equation}
that solved for $\dd T_d/\dd t$ gives the differential equation for the grain temperature
\begin{equation}
	\frac{\dd T_{d,i}}{\dd t} = A\frac{\dd T_g}{\dd t} \left(A+\int_0^\infty\frac{\dd B(E,T_{d,i})}{\dd T_{d,i}}Q_i(E)\dd E\right)^{-1}\,,
\end{equation}
where $A=2fn_g v_g k_B$.

This makes it possible to advance the solution of the solver by $T_{d,i}(t_0+\Delta t)=T_{d,i}(t_0) + \dd T_{d,i}/\dd t$ where $T_{d,i}(t_0)$ is determined \emph{only once at the beginning of the solver call} with the bisection method described e.g. in \cite{Dopcke2011}. As expected, this approach gives the same result as the bisection method, but with better solver stability and efficiency.

\section{Dust temperature convergence algorithm} \label{sect:convergence_dust}
The algorithm employed to compute the dust temperatures $T_{d,i}$ in the optically thick case is simple, but we report it here for the sake of clarity:
\begin{enumerate}[label=\arabic*.]
	\item initial guess on $\mathbf{T_d}$ ($T_{d,i}=T_{CMB}\,\forall i)$
	\item store $\mathbf{T_d}^{old}=\mathbf{T_d}$
	\item solving \eqn{eqn:beta_balance} $\forall i$ using bisection method
	\item update $\mathbf{T_d}$ with the $T_{d,i}$ values found
	\item if $ERR[\mathbf{T_d}^{old},\mathbf{T_d}]>\epsilon$ go to 2
	\item convergence found, $\mathbf{T_d}$ are the dust temperatures
\end{enumerate}
where $\epsilon=0.1$~K is the maximum error allowed and
\begin{equation}
 ERR[\mathbf{T_d}^{old},\mathbf{T_d}] = \max\left(|\mathbf{T_{d}^{old}} - \mathbf{T_{d}}|\right)\,,
\end{equation}
being $\mathbf{T_d}=\{T_{d,i}\,\forall i\}$ and analogously for $\mathbf{T_d}^{old}$.
The method described here is not necessary stable and convergent for any arbitrary case, but for all the models presented in this work it reaches quickly a convergence.

\section{Consistent interpolation} \label{appendix:advection}
Application~3 includes species which contain different atoms (e.g.~H$_2$O). In this application \krome is used in conjunction with the Adaptive Mesh Refinement code \ramses and molecular and atomic species are advected with passive scalars. When cells are refined or when solving the MHD equations the passive scalars are interpolated in space. The interpolation scheme applies slope limiters and the interpolation weights can be different for different passive scalars. If this scheme is applied na{\"i}vely the abundance of different atomic species, the metallicity, is therefore not always conserved \citep{Plewa1999}. Even though in each interpolation the error can be very small, over many time updates and / or repeated cell refinements the errors accumulate and become catastrophic.

To ensure conservation we apply a rescaling after doing the interpolation to either face values, for the MHD solver, or new cell centers, when refining cells. This corrects the interpolated abundances and conserves the metallicity. If we assume a known reference mass abundance $B_i$ for atom $i$ then the problem consists in finding rescaling factors $\omega_k$ such that
\begin{equation}
	\sum_k \omega_k \frac{c_{ik} m_i}{m_k} x_k = B_i\,,
\end{equation}
where $x_k$ is the interpolated mass abundance of species $k$, $m_k$ is the mass of species $k$, $m_i$ is the mass of atom $i$, and $c_{ik}$ is the multiplicity of the \ith atom in the \kth species (e.g.~$c_{\rm H, \rm H2O}=2$). In general there will be more species than atoms, and the choice of method to ensure conservation is not unique. We have to determine (in the space of the chemical abundances) the components of $\vec x_0 /\vec x$, where $\vec x$ are the uncorrected interpolated abundances and $\vec x_0$ are the real (correct, unknown) values. In principle, the distance for each component depends on the considered species (i.e.~the interpolation error on CO could be different from the error on H). To make the problem tractable and lower the degrees of freedom, we use a single rescaling factor $r_i$ for each type of atom. The rescaling factor for the \kth species is then constructed as the mass weighted average from the individual atomic rescaling factors
\begin{equation}
	\omega_k = \sum_j \frac{c_{jk} m_j}{m_k} r_j\,.
\end{equation}

Inserting this expression of $\omega_k$ above we get a linear set of equations
\begin{equation}
	\sum_j \left[ \sum_k \frac{c_{ik} c_{jk} m_i m_j}{m_k^2} x_k \right] r_j = B_i \,,
\end{equation}
which can be solved to find $r_j$. If the input mass abundances $x_k$ are positive, and the reference metallicities $B_i$ are all positive, then by construction the rescaling factors will be positive, resulting in a physically meaningful solution. For the \kth species with mass abundance $x_k$ we then apply the rescaling factor as
\begin{equation}
	x_k' = \left[ \sum_j \frac{c_{jk} m_j}{m_k}  r_j \right] x_k \,.
\end{equation}

Finally, the electron abundance is recomputed under the assumption of charge neutrality. The above procedure is very similar to what is used by \citet{Glover2010}. For a polynomial interpolator, as long as no slope limiting is applied, there will be no error in the interpolated metallicity, and no rescaling is needed. The problem arises when slope limiting is applied to some of the species due to e.g.~chemical gradients, and it is exactly in this case the rescaling procedure may break the slope limiting, and create spurious oscillations, though in practice we have not seen any indication that this is the case. We also note that in principle, if information about the quality of the interpolation of each species was available, this could be incorporated in to the above formula as weighting factors.

The problem of conservation does not affect \krome itself, and is related to the framework code employed, in our case \ramsess, but the support routines to establish $B_i$ and solve for $r_j$ are publicly available in \kromes. In Application~2, implemented with \enzos, only primordial chemistry is considered with species containing a single type of atom. Conservation is then trivially ensured by simply counting the total number of nuclei, for example of hydrogen, and then apply a renormalisation across all species containing that atom.

\bsp


\end{document}